\documentclass[10pt,conference,balance,letterpaper]{IEEEtran}
\usepackage{latexsym}
\usepackage{amsfonts}
\usepackage{amsmath}
\usepackage{amssymb}
\usepackage{color}
\usepackage{epsfig}
\usepackage{xspace}
\usepackage{graphicx}
\usepackage{subfigure}
 \usepackage{balance}
\usepackage{cite}
\usepackage[english]{babel}
\newcommand{\myhrule}{\rule[.5pt]{\hsize}{.5pt}}

\newcommand{\eat}[1]{}

\newcommand{\sstab}{\rule{0pt}{8pt}\\[-2.4ex]}

\newcommand{\bi}{\begin{itemize}}
\newcommand{\ei}{\end{itemize}}

\newcommand{\mat}[2]{{\begin{tabbing}\hspace{#1}\=\+\kill #2\end{tabbing}}}

\newcommand{\be}{\begin{enumerate}}
\newcommand{\ee}{\end{enumerate}}
\newcommand{\beqn}{\begin{eqnarray*}}
\newcommand{\eeqn}{\end{eqnarray*}}

\newcommand{\stitle}[1]{\vspace{1ex}\noindent{\bf #1}}
\newcommand{\etitle}[1]{\vspace{0.5ex}\noindent{\em \underline{#1}}}

\newcommand{\ie}{\emph{i.e.,}\xspace}
\newcommand{\eg}{\emph{e.g.,}\xspace}
\newcommand{\wrt}{\emph{w.r.t.}\xspace}
\newcommand{\aka}{\emph{a.k.a.}\xspace}

\newcommand{\If}{\mbox{\bf if}\ }
\newcommand{\Let}{\mbox{\bf let}\ }

\newcommand{\Then}{\mbox{\bf then}\ }

\newcommand{\While}{\mbox{\bf while}\ }

\newcommand{\Do}{\mbox{\bf do}\ }

\newcommand{\For}{\mbox{\bf for}\ }

\newcommand{\Not}{\mbox{\bf not}\ }

\newcommand{\Return}{\mbox{\bf return}\ }

\newcommand{\kw}[1]{{\ensuremath {\mathsf{#1}}}\xspace}

\newcounter{ccc}
\newcommand{\bcc}{\setcounter{ccc}{1}\theccc.}
\newcommand{\icc}{\addtocounter{ccc}{1}\theccc.}

\newcommand{\NP}{{\sc np}\xspace}

\newcommand{\G}{{\cal G}}

\newcommand{\V}{{\cal V}}
\newcommand{\E}{{\cal E}}
\newcommand{\eop}{\hspace*{\fill}\mbox{$\Box$}}     
\newcounter{example}
\renewcommand{\theexample}{\arabic{example}}
\newenvironment{example}{
        \vspace{1.5ex}
        \refstepcounter{example}
        {\noindent\bf Example \theexample:}}{
        \eop\vspace{1.5ex}}

\renewcommand{\ni}{\noindent}

\newcommand{\nthesection}{\arabic{section}}
\newcounter{theorem}
\renewcommand{\thetheorem}{\arabic{theorem}}
\newcounter{prop}
\renewcommand{\theprop}{\arabic{theorem}}
\newcounter{property}
\renewcommand{\theprop}{\arabic{theorem}}
\newcounter{lemma}
\renewcommand{\thelemma}{\arabic{theorem}}
\newcounter{cor}
\renewcommand{\thecor}{\arabic{theorem}}
\newenvironment{theorem}{\begin{em}
        \refstepcounter{theorem}
        {\vspace{1ex} \noindent\bf  Theorem  \thetheorem:}}{
        \end{em}\eop\vspace{1ex}} 
\newenvironment{prop}{\begin{em}
        \refstepcounter{theorem}
        {\vspace{1ex}\noindent \bf Proposition \theprop:}}{
        \end{em}\eop\vspace{1ex}}

\newenvironment{cor}{\begin{em}
        \refstepcounter{theorem}
        {\vspace{1ex}\noindent\bf Corollary \thecor:}}{
        \end{em}\eop\vspace{1ex}} 
\newcounter{definition}[section]
\renewcommand{\thedefinition}{\nthesection.\arabic{definition}}

\newcounter{alg}[section]
\renewcommand{\thealg}{\nthesection.\arabic{alg}}

\newcounter{arule}
\renewcommand{\thearule}{\arabic{arule}}

\newcounter{claim}
\renewcommand{\theclaim}{\arabic{claim}}

\renewcommand{\texttt}[1]{{\small\textsf{#1}}}

\newcommand{\dist}{\kw{dist}}
\newcommand{\distV}{\kw{distVec}}

\newcommand{\refree}{{\sc ref}\xspace}

\newcommand{\cc}{{\sc cc}\xspace}

\newcommand{\ccs}{{\sc cc}s\xspace}
\newcommand{\bc}{{\sc bcc}\xspace}

\newcommand{\bccs}{{\sc bcc}s\xspace}

\newcommand{\dra}{{\sc dra}\xspace}
\newcommand{\dras}{{\sc dra}s\xspace}
\newcommand{\lcover}{{\sc lmc}\xspace}
\newcommand{\scover}{{\sc sc}\xspace}
\newcommand{\vcover}{{\sc vc}\xspace}
\newcommand{\bcsketch}{{\sc bc-Sketch}\xspace}
\newcommand{\super}{\textsc{super}\xspace}

\newcommand{\gdp}{{\sc bgp}\xspace}
\newcommand{\spaceL}{\kw{space_L}\xspace}
\newcommand{\spaceN}{\kw{space_N}\xspace}

\newcommand{\timec}{\kw{time}\xspace}
\newcommand{\sizec}{\kw{size}\xspace}

\newcommand{\compDRAs}{\kw{compDRAs}\xspace}
\newcommand{\metis}{{\sc Metis}\xspace}
\newcommand{\disland}{{\sc disLand}\xspace}
\newcommand{\bisearch}{{\sc BSearch}\xspace}
\newcommand{\ch}{{\sc ch}\xspace}
\newcommand{\arcflag}{{\sc arcFlag}\xspace}

\title{Distance Landmarks Revisited for Road Graphs}

\author{
Shuai Ma{\small $^{1}$}\hspace{5ex} Kaiyu Feng{\small $^{1}$}\hspace{5ex} Haixun Wang{\small $^{2}$}\hspace{5ex} Jianxin Li{\small $^{1}$}\hspace{5ex} Jinpeng Huai{\small $^{1}$}\\
\vspace{0.2ex}
\begin{tabular}{cc}
\hspace{-6ex}{\small $^{1}$ SKLSDE Lab, Beihang University, China}  &\hspace{4ex}{\small $^{2}$ Google Research, USA}\\
\end{tabular}\\
{\small \{mashuai, fengky, lijx, huaijp\}@buaa.edu.cn} \hspace{2ex} {\small haixun@google.com}
}

\date{}

\begin{document}
\maketitle

\begin{abstract}
Computing shortest distances is one of the fundamental problems on graphs, and remains a {\em challenging} task today.
{\em Distance} {\em landmarks} have been recently studied for shortest distance queries with an auxiliary data structure, referred to as {\em landmark} {\em covers}.
This paper studies how to apply distance landmarks for fast {\em exact} shortest distance query answering on large road graphs.
However, the {\em direct} application of distance landmarks is {\em impractical} due to the high space and time cost.
%
To rectify this problem, we investigate novel techniques that can be seamlessly combined with distance landmarks.
We first propose a notion of {\em hybrid landmark covers}, a revision of landmark covers.
Second, we propose a notion of {\em agents}, each of which represents a small subgraph and  holds good properties for fast distance query answering.
We also show that agents can be computed in {\em linear time}.
Third, we introduce graph partitions to deal with the remaining subgraph that cannot be captured by agents.
Fourth, we develop a unified framework that seamlessly integrates our proposed techniques and existing optimization techniques, for fast shortest distance query answering.
Finally, we experimentally verify that our techniques significantly
improve the efficiency of shortest distance queries, using real-life road graphs.
\end{abstract}

\section{Introduction}
\label{sec-intro}

We study the {\em node-to-node shortest distance} problem on large graphs: given a weighted undirected graph $G(V, E)$ with non-negative edge weights and two nodes of $G$, the source $s$ and the target $t$, find the shortest distance from $s$ to $t$ in $G$. We allow the usage of auxiliary structures  generated by preprocessing, but restrict them to have a moderate size (compared with the input graph).  In this work, we are only interested in {\em exact} shortest distances on {\em large} graphs.

Finding shortest distances, a twin problem of {\em finding shortest paths}, is one of the fundamental problems on graphs, and has found its usage as a building block in various applications, \eg measuring the closeness of nodes in social networks and Web graphs~\cite{LappasLT09,PotamiasBCG09,SarmaGNP10}, and finding the distances between physical locations in road networks~\cite{WuXDCZZ12}.

Algorithms for shortest distances have been studied since 1950's and still remain an {\em active} area of research. The classical one is Dijkstra's algorithm~\cite{CormenLRS01} due to Edsger Dijkstra. Dijkstra's original algorithm runs in $O(n^2)$~\cite{Dijkstra59}, and the enhanced implementation with Fibonacci heaps runs in $O(n\log n + m)$ due to Fredman \& Tarjan~\cite{FredmanT84}, where $n$ and $m$ denote the numbers of nodes and edges in a graph, respectively. The latter remains asymptotically the fastest known solution on arbitrary undirected graphs with non-negative edge weights~\cite{ThorupZ05}.

However, computing shortest distances remains a challenging problem, in terms of both time and space cost, for large-scale graphs such as Web graphs, social networks and road networks. The Dijkstra's algorithm~\cite{FredmanT84} is not acceptable on large graphs (\eg with tens of millions of nodes and edges) for online applications ~\cite{PotamiasBCG09}. Therefore, a lot of optimization techniques have been recently developed to speed up the computation~\cite{PotamiasBCG09,SarmaGNP10,WuXDCZZ12,LubyR89,GeisbergerSSD08,Wei10,SankaranarayananSA09,SandersS05,ChengKCC12}.

{\em Distance landmarks} (\aka\ {\em distance oracles}, see Section \ref{subsec-dislandmarkdef} for details) are data structures that support efficient shortest distance query answering, and have been recently studied in both theory~\cite{ThorupZ05,MozesS12} and practice~\cite{PotamiasBCG09,SarmaGNP10,SankaranarayananS10}. An $n\times n$ {\em triangular matrix} of size $n^2/2$ for all-pair shortest distances can be computed in $O(n^2\log n + $ $mn)$ time, using Dijkstra's algorithm~\cite{FredmanT84}, where $n$ and $m$ are the numbers of nodes and edges, respectively. With the distance matrix, shortest distance queries can be answered in $O(1)$ time.
This solution, however, is {\em not practical} on large graphs: the preprocessing time is too long, and even if one is willing to wait that long, the matrix is too large to be stored effectively. For instance, the matrix of a graph with one million nodes needs about $1,862$ GB memory (here the distance entries are stored as $4$-byte integers).

Distance landmarks aim at {\em striking a balance} between the efficiency benefits of answering shortest distance queries and the time and space cost of computing and storing them. And distance landmarks have already been adopted for answering {\em approximate} shortest distances~\cite{PotamiasBCG09,SarmaGNP10,ThorupZ05,SankaranarayananS10}, and for answering {\em exact} shortest distances on directed graphs~\cite{GoldbergH05,MozesS12}. However, how to apply distance landmarks for answering exact shortest distances on undirected graphs is mainly limited to pure theoretical analyses~\cite{ThorupZ05}.

\eat{

In this work, we show, both theoretically and experimentally, that distance landmarks can not be directly use for real-life large graph. Nevertheless, we propose a set of techniques for distance landmarks, which together work well for answering shortest distance queries on large graphs.
}

\stitle{Contributions \& Roadmap}. To our knowledge, we are among the first to study the application of distance landmarks for fast exact shortest distance queries on large undirected graphs.

\sstab (1) We develop an approximation algorithm with a constant  factor $2$ to analyze distance landmarks by establishing connections with vertex covers (Section~\ref{sec-analysis}), based on which we show that the {\em direct} application of distance landmarks is not practical for large-scale graphs.
We then propose {\em hybrid landmark covers}, a revised notion of traditional landmark covers, to reduce the space cost (Section~\ref{sec-analysis}).

\sstab (2) We propose a notion of {\em agents} such that each agent represents a small subgraph, referred to as  {\em deterministic routing areas} (\dras)  (Section~\ref{sec-agent}). Then landmarks are only built for agents, instead of the entire graph. Hence, both space and time cost are reduced. We give an analysis of agents and \dras, based on which we develop a linear time algorithm for computing \dras along with their maximal agents.  As shown in the experimental study, on average about 1/3 nodes of a graph are captured by agents and their \dras.

\sstab (3) We introduce the {\em bounded graph partitioning} problem (\gdp) to deal with the remaining subgraph that cannot be captured by the \dras of agents, and show that the problem is \NP-complete (Section~\ref{sec-decomposition}). We then propose a notion of {\em \super} graphs that combine graph partitions with hybrid landmark covers to support efficient shortest distance answering. We also build connections between the traditional graph partitioning problem and the \gdp problem, and utilize the traditional graph partitioning approaches, \eg\ \metis, to solve the problem.
As shown by the experiments, \metis works well.

\sstab (4) We propose a unified framework \disland for fast shortest distance query answering (Section~\ref{sec-query}), which seamlessly combines distance landmarks with agents, graphs partitions (\super graphs), and existing speed-up techniques~\cite{WuXDCZZ12,WagnerW07}.

\sstab (5) Using real-life large road graphs, we conduct an extensive experimental study (Section~\ref{sec-expt}).
We find that our \disland scales well with large graphs, \eg it takes $0.28 \times 10^{-3}$ seconds on graphs with $2.4$ $\times$ $10^7$ nodes and $5.7$ $\times$ $10^7$ edges. Moreover, \disland is $9.4$, $134.9$, and $14,540.1$ times faster than \ch~\cite{GeisbergerSSD08}, \arcflag~\cite{MohringSSWW05}, and bidirectional Dijkstra~\cite{LubyR89}, respectively. Moreover, the auxiliary structures occupy only a moderate size of space (about $1/2$ of the input graphs), and can be pre-computed efficiently.

\vspace{0.5ex}
Due to the space constraint, we defer all the proofs to~\cite{full}.

\vspace{-0.5ex}
\stitle{Related work}.
(1) Algorithms for node-to-node shortest distances have been extensively studied since 1950's, and fall into different categories
 in terms of different criteria:

 \bi
 \item exact distances~\cite{WuXDCZZ12,Dijkstra59,FredmanT84,LubyR89,GeisbergerSSD08,SankaranarayananSA09,SandersS05,GoldbergH05,MozesS12,ChengKCC12,MozesS12,ChanL07,SaundersT07,WagnerW07}
     and approximate distances~\cite{PotamiasBCG09,SarmaGNP10,ThorupZ05,SankaranarayananS10};

\vspace{0ex}
 \item memory-based~\cite{PotamiasBCG09,SarmaGNP10,WuXDCZZ12,Dijkstra59,FredmanT84,LubyR89,GeisbergerSSD08,Wei10,SankaranarayananSA09,SandersS05,
ThorupZ05,MozesS12,SankaranarayananS10,SaundersT07,WagnerW07} and disk-based algorithms~\cite{ChengKCC12,ChanL07};

\vspace{0ex}
 \item for unweighted~\cite{PotamiasBCG09,SarmaGNP10,Wei10} and weighted graphs~\cite{WuXDCZZ12,Dijkstra59,FredmanT84,LubyR89,GeisbergerSSD08,SankaranarayananSA09,GoldbergH05,MozesS12,SandersS05,ChengKCC12,ThorupZ05,MozesS12,SankaranarayananS10,ChanL07,SaundersT07,WagnerW07}; and

\vspace{0ex}
 \item for directed~\cite{SaundersT07,GoldbergH05,MozesS12} and undirected graphs~\cite{PotamiasBCG09,SarmaGNP10,WuXDCZZ12,Dijkstra59,FredmanT84,LubyR89,GeisbergerSSD08,Wei10,SankaranarayananSA09,SandersS05,ChengKCC12,ThorupZ05,MozesS12,SankaranarayananS10,ChanL07,WagnerW07}.
 \ei

In this work, we study the memory-based exact shortest distance problem on weighted undirected large real-world graphs.
None of the previous work has {\em experimentally} studied how to apply distance landmarks for solving this problem.

\sstab (2)  Distance landmarks have been recently investigated for {\em approximate} shortest distance queries~\cite{ThorupZ05,PotamiasBCG09,SarmaGNP10,SankaranarayananS10}, and for answering {\em exact} shortest distances on directed graphs~\cite{GoldbergH05,MozesS12}. However, how to apply distance landmarks for answering exact shortest distances on undirected graphs is mainly limited to pure theoretical analyses~\cite{ThorupZ05}.
Nevertheless, in this work, we investigate how to utilize distance landmarks to speed-up shortest distance queries on real-life large road graphs.

\sstab(3) There has recently been extensive work on speed-up techniques for shortest distance queries: bidirectional search~\cite{LubyR89},
hierarchical approaches~\cite{GeisbergerSSD08}, node and edge labeling~\cite{MohringSSWW05,SankaranarayananSA09} and shortcuts~\cite{SandersS05} (see~\cite{WuXDCZZ12,WagnerW07} for two recent surveys). These techniques are complementary to our work, and can be incorporated into our approach.
We have indeed seamlessly integrated the \ch \cite{GeisbergerSSD08} and \arcflag \cite{MohringSSWW05} techniques with distance landmarks into our framework.

\sstab(4) Graph partitioning has been extensively studied since 1970's~\cite{kl70,Karypis98,YangYZK12},
and has been used in various applications, \eg circuit placement, parallel computing and scientific simulation~\cite{YangYZK12}.
The graph partitioning problem considered in this work differs from the traditional one that it concerns more on the number of nodes with edges across different partitions, instead of the number of edges with endpoints across  different partitions. Nevertheless,
we build connections between these two problems, and make use of the existing approaches, \eg~\metis~\cite{Karypis98}, to solve the graph partitioning problem considered in this work. It is also worth mentioning that graph partitioning has already been used to speed-up Dijkstra's algorithm~\cite{MohringSSWW05}.

\sstab(5) Agents and deterministic routing areas proposed in this study (Section~\ref{sec-agent}) are significantly different (from definitions to analyses to algorithms) from the 1-dominator sets proposed in~\cite{SaundersT07}. Moreover, the latter are for shortest path queries on nearly acyclic directed graphs, which is not appropriate for real-life large graphs, as these graphs typically contain a large strongly connected components~\cite{BroderKMRRSTW00}.

\section{Preliminary}
\label{sec-preliminary}

In this section, we first present basic notations of graphs. We then introduce the notion of distance landmarks.

\subsection{Graph Notions}
\label{subsec-graphdefs}

We first introduce graphs and the related concepts.

\etitle{Graphs}. A {\em weighted undirected graph} (or simply a {\em graph})
is defined as $G (V$, $E$, $w)$, where
(1) $V$ is a finite set of nodes;
(2) $E\subseteq V \times V$ is a finite set of edges, in which $(u, v)$ or $(v, u)$ $\in E$ denotes an undirected edge between nodes $u$ and $v$; and
(3) $w$ is a total weight function that maps each edge in $E$ to a positive rational number.

We simply denote $G (V$, $E$, $w)$ as  $G(V, E)$ when it is clear from the context.

\etitle{Subgraphs}. Graph $H(V_s, E_s,  w_{s})$ is a {\em subgraph} of graph
$G(V$, $E,  w)$ if (1) for each node
$u\in V_s$, $u\in V$, and, moreover, (2) for each edge $e\in E_s$, $e\in E$ and $w_{s}(e)$ = $w(e)$.
That is, $H$ contains a subset of nodes and a subset of edges of $G$.

We also denote subgraph $H$ as $G[V_s]$ if $E_s$ is {\em exactly} the set of edges appearing in $G$ over $V_s$.

\etitle{Paths and cycles}.
A {\em simple path} (or simply a {\em path}) $\rho$ is
a sequence of nodes $v_1/\ldots/v_n$ with no repeated nodes, and, moreover, for each $i\in[1, n-1]$, $(v_i$, $v_{i+1})$ is an edge in $G$.

A {\em simple cycle} (or simply a {\em cycle}) $\rho$ is
a sequence of nodes $v_1/\ldots/v_n$ with $v_1 = v_n$ and no other repeated nodes, and, moreover, for each $i\in[1, n-1]$, $(v_i$, $v_{i+1})$ is an edge in $G$.

The {\em length} of a path or cycle $\rho$ is
the sum of the weights of its constituent edges, \ie $\sum_{i=1}^{n-1} w(v_i, v_{i+1})$.

We say that $v_{i+1}$ (resp. $v_i$) is a {\em neighbor} of $v_i$ (resp. $v_{i+1}$).

We also say that a node is {\em reachable} to another one if there exists a path between these two nodes.

\etitle{Shortest paths and distances}.
A {\em shortest path} from one node $u$ to another node $v$ is a path whose length is minimum among all the paths from $u$ to $v$.

The {\em shortest distance} between nodes $u$ and $v$, denoted by $\dist(u, v)$, is the length of a shortest path from $u$ to $v$.

\etitle{Connected components}.
A {\em connected component} (or simply a \cc) of a graph is a subgraph in which any two nodes are connected by a path, and is connected to no additional nodes.
A graph is connected if it has exactly one connected component, consisting of the entire graph.

\etitle{Cut-nodes and bi-connected components}.
A {\em cut-node} of a graph is a node whose removal increases the number of connected components in the graph.

A {\em bi-connected component} (or simply a \bc) of a graph is a subgraph consisting of a maximal set of edges such that any two edges in the set must lie on a common simple cycle.

\eat{
\begin{figure}[tb!]
\begin{center}
\includegraphics[scale=0.58]{./landmarks.eps}
\end{center}
\vspace{-5ex}
\caption{An example landmark cover}
\label{fig-landmarkCover}
\vspace{-4ex}
\end{figure}
}

\subsection{Distance Landmarks}
\label{subsec-dislandmarkdef}

We next introduce the notion of distance landmarks \cite{PotamiasBCG09}.

Consider an ordered set of $l$ vertices $D$ = $<x_{1},\dots, x_{l}>$ such that for each $i\in[1,l]$, $x_i$ is a distinct node in graph $G$.

We say that $D$ is a {\em landmark cover} of graph $G$ if and only if for any node pair $(u, v)$ in $G$ with $u$ reachable to $v$, there exists a {\em landmark} $x_i$ ($1\le i\le l$) in $D$ such that the shortest distance $\dist(u,v)$ = $\dist(u, x_i)+\dist(x_i, v)$.
This is achieved by representing each node in $G$ as a vector of shortest distances to the set of landmarks in $D$.
More specifically,  each node $u \in V$ is represented as an $l$-dimensional vector $\distV(u)$:

\hspace{2ex}$\distV(u)$ = $<\dist(x_1, u), \dots, \dist(x_{l},u)>$.


The \lcover problem  is to find a landmark cover with a minimum number of landmarks in a graph. The problem is unfortunately intractable, as shown below.

\begin{prop}
The \lcover problem is \NP-complete~\cite{PotamiasBCG09}.
\end{prop}

To reduce its computational complexity, an $O(\log n)$-approximation algorithm was proposed by using the approximation algorithms for the {\em set cover} (\scover) problem~\cite{PotamiasBCG09}. This  algorithm, however, runs in cubic time, and cannot be directly used for large graphs, as already been observed in~\cite{PotamiasBCG09}.

\eat{We next briefly present the solution developed in~\cite{PotamiasBCG09}. Consider  graph $G(V, E)$. Let $U = V\times V$ be a set of elements, \ie the set of node pairs, and $C = \{S_u \ |\ u\in V\}$ be a collection of subsets in $U$ such that for each $u \in V$,  $S_u\in C$ is the set of node pairs $(s, t)$ such that $u$ lies on a shortest path from $s$ to $t$.  Then the algorithm for the \scover problem can be used to solve the \lcover problem, from which an $O(\log n)$-approximation algorithm was obtained.}


\stitle{Remarks}. (1) With a landmark cover $D$, the exact shortest distance $\dist(u,v)$ for any node pair $(u, v)$ can be computed in $O(|D|)$ time, where $|D|$ is the number of landmarks in $D$. This is obvious as $\dist(u,v)$ = $\kw{min}\{\dist(u, x_i)$ + $\dist(x_i, v)\ | \ x_i\in D \}$.
(2) As a landmark cover $D$ occupies $|D|$ $(|V|-1)$ space, its size $|D|$ must be small in order to apply it on large graphs.

\section{Distance Landmarks Revisited}
\label{sec-analysis}
In this section, we first show that it is not practical to {\em directly} utilize landmark covers due to the high space cost.
We then propose a notion of {\em hybrid landmark covers} to alleviate this problem.
Here we consider a graph $G(V, E, w)$.

\subsection{Landmark Covers}

To give a more accurate estimation of landmark covers, we develop an approximation algorithm with a constant factor $2$.
Recall that the \scover based algorithm (Section~\ref{subsec-dislandmarkdef}, \cite{PotamiasBCG09}) has an approximation factor of $O(\log n)$.
To do this, we first present a notion of redundant-edge-free (\refree) graphs. We then build the relationship between the \lcover problem and the clasical {\em vectex cover} (\vcover) problem on \refree graphs, which leads to a $2$-approximation algorithm. Finally, we evaluate the cost of landmark covers with the approximation algorithm.

A {\em vertex cover} of a graph is a set of nodes such that each edge of the graph is incident to at least one node of the set.
The \vcover problem is to find a minimum set of vertex covers, a classical optimization problem known to be \NP-complete~\cite{GaJo79}.

\begin{table*}[th]
\vspace{-1.5ex}
\label{tab-landmark-overhead}
\caption{Overhead of landmark covers vs. original graphs}
\vspace{-2ex}
\begin{center}
\begin{scriptsize}
\vspace{-2ex}
\begin{tabular}{|c|r||r|c||r|c||r|}
\hline
\multicolumn{2}{|c||} {Graphs $G(V, E)$}  &  \multicolumn{4}{c||} {Landmark covers $D$} & \\
\cline{1-6}
name                     & size\hspace{0.5ex}({\sc mb})    & $\le|D|\le$ & $\le\frac{|D|}{|V|}\le$ (\%)  & $\le$ size $\le$\hspace{0.5ex}({\sc gb})& $\le\frac{\sizec(D)}{\sizec(G)}\le$ & $\timec$ (s) \\

\hline\hline
\hspace{-1ex}CO\hspace{-1.5ex}       &    9.62      & [181,276,\ \   362,552]   & [41.6, 83.2]         & [588.42,\ \   1,176.83] & [$6.27\times 10^4$,\ \   $1.25\times 10^5$]      & 34.1\\
\hline
\hspace{-1ex}FL\hspace{-1.5ex}       &   24.59       & [447,486,\ \   894,972]   & [41.8, 83.6]          & [3,568.67,\ \    7,137.33] & [$1.49\times 10^5$,\ \   $2.97\times 10^5$]      & 391.8\\
\hline
\hspace{-1ex}CA\hspace{-1.5ex}       &   42.54       & [761,662,\ \   1,523,324] & [40.3, 80.6]     & [10,730.05,\ \   21,460.09]  & [$2.58\times 10^5$,\ \   $5.17\times 10^5$]    & 1,205.6\\
\hline
\hspace{-1ex}E-US\hspace{-1.5ex}     &   80.17       & [1,450,115,\ \   2,900,230] & [40.3, 80.6]     & [38,880.24,\ \    77,760.48] & [$4.97\times 10^5$,\ \   $9.93\times 10^5$]     & 4,315.3\\
\hline
\hspace{-1ex}W-US\hspace{-1.5ex}     &   139.24       & [2,545,995,\ \   5,091,990] & [40.7, 81.3]     & [118,786.74,\ \   237,573.47] & [$8.74\times 10^5$,\ \   $1.75\times 10^6$]    & 12,984.3\\
\hline
\hspace{-1ex}C-US\hspace{-1.5ex}     &   312.10       & [5,811,428,\ \   1,1622,856] & [41.3, 82.5]     & [609,721.69,\ \   1,219,443.38] & [$2.00\times 10^6$,\ \   $4.00\times 10^6$]  & 66,996.9\\
\hline
\hspace{-1ex}US\hspace{-1.5ex}       &   531.63      & [9,737,381,\ \ 19,474,762]  & [40.7, 81.3]      &[1,737,359.48, 3,474,718.95]   &  [$3.35\times 10^6$,\ \ $6.69\times 10^6$] & 196,194.6 \\
\hline
\end{tabular}
\vspace{-4.5ex}
\end{scriptsize}
\end{center}
\vspace{-2ex}
\end{table*}

Graphs often contain redundant edges when distance queries are concerned.
Graph $G$ is {\em redundant-edge-free} (\refree) if it contains no {\em redundant} edges, where
an edge $(u,v)$ is redundant if its removal has no effects on the shortest distance $\dist(u, v)$.

By the definition of \refree graphs above, it is trivial to see that \refree graphs preserve shortest distances, and that a graph may have multiple \refree graphs.
We next build the relationship between landmark covers and vertex covers, stated as follows.

\begin{theorem}
\label{thm-equiv-vclc}
For any \refree graph $G$, a set $S$ of nodes is a landmark cover of $G$ iff $S$ is a vertex cover of $G$.
\end{theorem}

As a consequence, the \lcover problem is identical to the \vcover problem on \refree graphs.

\begin{figure}[t]
\begin{center}
{\small
\begin{minipage}{3.36in}
\myhrule \vspace{-2ex}
\mat{0ex}{
\sstab {\sl Input:\/} \= A weighted undirected graph $G(V, E, w)$.\\
{\sl Output:\/} A landmark cover $D$ of $G$. \\

\sstab\bcc\ \= Remove redundant edges from $G$; \\
\icc\> Compute a vertex cover $D$ of $G$;\\
\icc\> \Return $D$.
}
\vspace{-3ex}
\myhrule
\end{minipage}
}
\end{center}
\vspace{-2ex}
\caption{$2$-approximation algorithm for computing landmark covers}
\label{alg-landmark-cover}
\vspace{-4ex}
\end{figure}

\etitle{Approximation algorithm}. It is well-known that the \vcover problem has a 2-approximation algorithm~\cite{approx03}, which basically computes a {\em maximal matching} of a graph by greedily picking edges and removing all endpoints of the picked edges~\cite{CormenLRS01}.
Following from Theorem~\ref{thm-equiv-vclc}, we obtain a $2$-approximation algorithm for the \lcover problem, presented in Fig.~\ref{alg-landmark-cover}.

Given a graph $G(V, E)$, the algorithm first computes an \refree graph of $G$ by removing redundant edges (line 1). It then computes a vertex cover $D$ of the \refree graph (line 2), and simply returns $D$ as a landmark cover of $G$ (line 3).

Note that testing whether an edge $(u, v)$ is redundant in a graph $G(V, E, w)$ is typically efficient. When computing  $\dist(u, v)$ using Dijkstra's algorithm on graph $G(V, E\setminus\{(u, v)\}, w)$, if $\dist(u, v') > w(u, v)$ for any node $v'$ before reaching $v$, it is easy to verify  that $(u, v)$ is not a redundant edge. Moreover, for a large portion of edges $(u, v)$, its weight $w(u, v)$ is exactly the shortest distance $\dist(u, v)$ in real-life graphs such as road networks. Hence, our \vcover based algorithm is typically much faster than the \scover based algorithm~\cite{PotamiasBCG09}, though they have the same time complexity.

\vspace{-.5ex}
\stitle{Remarks}. The $2$-approximation algorithm allows us to have both lower and upper bounds for the sizes and space cost of landmark covers.
If the algorithm returns a landmark cover $D$, then the lower and upper bounds for the size of the {\em optimal landmark cover} are $|D|/2$ and $|D|$, respectively.

\vspace{-.5ex}
\stitle{Findings on landmark covers}. We next experimentally test the overhead of landmark covers with our approximation algorithm.
We tested seven real-life datasets from \cite{dimacs-datasets} (please refer to Section~\ref{sec-expt} for details about the datasets and experimental settings). We adopted the adjacency-list representation \cite{CormenLRS01} for graphs when counting their space cost, and assumed that nodes and distances were stored as $4$-byte integers.

The experimental results shown in Table~1 tell us that:

\sstab (1) The size of an optimal landmark cover is large, and typically 40\%--80\% of the nodes in a graph are landmarks.

\sstab (2) The space cost of a landmark cover is huge, and is typically more than $10^4$--$10^6$ times of the graph itself.
For instance, the landmark cover of the US graph with $1/2$ GB space may incur a space cost of more than $1.74\times 10^6$ GB.

\sstab (3) Computing landmark covers of large graphs is inefficient. It took our algorithm more than 2 days 6 hours on the US dataset.
It is worth mentioning that here we only compute the landmarks nodes, not including computing the shortest distances between graph nodes and landmarks.
Furthermore, the directly usage of \scover based algorithm~\cite{PotamiasBCG09} is even worse, due to its high space and time cost (it even runs out of memory  --16GB-- for the smallest CO dataset on our testing machine).

Hence, it suffices to conclude that the direct application of distance landmarks as~\cite{PotamiasBCG09} is impractical for large graphs.

\eat{
space comparison

6865.4545454545454545454545454545

Nodes space = number * 8 bytes
Edge  space = number * 2*8 bytes (Neighborhood storage)

Graph space

nodes               edges     =  G/1024*1024*1024

nodes space              edges space      landmark covers
392872              1936384         989722632*16
935360              4269280     5566630509*16
1556040             6877472     15126381576*16
3485328             16913056    79060998880*16
8563008             43404768    480196984375*16
15126520            74523872    1474120192308*16
28788984            140449824   5235235700758*16
50096832            243970336   16002422228217*16

112654528           548679936   81031326018060*16
191578776           933333504

G space   (G)                  D space

0.002169288694858551025390625           14.75
0.0048471987247467041015625             82.95012964308261871337890625
0.007854320108890533447265625           225.4007950611412525177001953125
0.01899747550487518310546875            1178.10393442213535308837890625
0.0483987629413604736328125             7155.50064389407634735107421875
0.083493433892726898193359375           21966.116692356765270233154296875
0.157615922391414642333984375           78011.117876611649990081787109375
0.2738713920116424560546875             238454.71977098286151885986328125
0.615915715694427490234375              1207460.95566403865814208984375
1.047656200826168060302734375
}

\subsection{Hybrid Landmark Covers}

The naive matrix approach stores the pre-computed all-pair shortest distances of a graph $G(V, E)$, and takes $|V|(|V| - 1)/2$ space. And the landmark approach was proposed to reduce the space cost to $|V||D|$, where $|D|$ is the size of a landmark cover.
One might believe that the landmark approach always incurs less space than the matrix approach. It is, however, not the case as shown by the following example.

\vspace{-0.5ex}
\begin{example}
\label{exm-revised-landmarks} Consider node $x$ in a landmark cover $D$ that lies on the shortest paths of a set $\{(u_1, u_2)$, $\ldots$, $(u_{2k-1}$, $u_{2k})\}$ of $k$ node pairs in a graph, where nodes $u_i \ne u_j$ for any $i\ne j$ $\in$ $[1, 2k]$. Then node $x$ takes $k$ space in the naive approach, by directly adding edges to connect those $k$ node pairs, while it takes $2k$ space in the landmark approach, by adding edges between $x$ and each of the $2k$ nodes.
\end{example}
\vspace{-0.5ex}

This motivates us to propose a {\em hybrid} approach combining the naive approach with the landmark one.
To do this, we first define the following notions.

Consider a node $x$ in a graph $G$.
Let $P_{x}$ be a set of node pairs such that $x$ lies on their shortest paths, and let $N_{x}$ be the set of distinct nodes in $P_{x}$.
For a landmark node $x$, we only store the shortest distances between $x$ and the node in $N_{x}$, instead of all the nodes in the graph as~\cite{PotamiasBCG09}.
Hence, the space cost of making $x$ a landmark, denoted by $\spaceL(x)$, is exactly $|N_{x}|$.
Alternatively, the naive approach incurs a space cost of $|P_x|$, denoted by $\spaceN(x)$, by storing the shortest distances for each node pair in $P_x$.

Consider an ordered set of $l$ vertices $D$ = $<x_{1},\dots, x_{l}>$ such that (a) for each $i\in[1,l]$, $x_i$ is a node in graph $G$,
and (b) $P_{x_i}\cap P_{x_j} = \emptyset$ for any $i\ne j\in[1, l]$.

\etitle{Hybrid landmark covers}.  We say that $\tilde{D}$ = $(D, E_{D}^-)$ is a {\em hybrid landmark cover} of graph $G$ if and only if:

\sstab(1) for each $x_i$ $(i\in [1, l])$, $\spaceL(x_i) \le \spaceN(x_i)$,

\sstab(2) there exist no other nodes $x$ in $G$, but $x \not\in D$, such that $\spaceL(x) \le \spaceN(x)$, and

\sstab(3) $E_{D}^-$ is a set edges, denoting all the node pairs of $G$ such that no landmarks in $D$ lie on their shortest paths.

We also call $E_{\tilde{D}}$  = $\{(u, x)\ |\ u\in N_{x}, x\in D\} \cup E_{D}^-$ the set of edges {\em enforced} by a hybrid landmark cover $\tilde{D}$.

\stitle{Remark}. (1) Essentially, $D$ consists of a maximal set of landmarks such that the space cost of each landmark in the set is not larger than the corresponding naive cost.

\sstab(2) A hybrid landmark cover $\tilde{D}$ of a graph can be treated as another graph with the same set of nodes, but with a different set of edges, \ie the set  $E_{\tilde{D}}$ of enforced edges. Similarly, the naive approach transforms a graph into a complete graph.
This provides a unified view for these two approaches.


\sstab(3) Computing hybrid landmark covers on large graphs remains very challenging. Indeed, they cannot be directly used in practice as well.
 As will be seen in Section~\ref{sec-decomposition}, we build hybrid landmark covers \wrt a (small) subset of nodes in graph $G$.
 In the following, we will explore techniques to support efficient shortest distance queries on large real-life road graphs.

\eat{\vspace{-0.5ex}
Hence, the most {\em challenging task} is to explore {\em novel} techniques that could be seamlessly incorporated with hybrid landmark covers to reduce the space and time cost, while maintaining the efficiency benefits of answering shortest distance queries on large graphs.

As will be seen in Section~\ref{sec-decomposition}, we will also build partial landmark covers \wrt a (small) subset of nodes in graph $G$ only, not for the entire $G$. In this way, both space and time cost are significantly reduced.
We will also experimentally verify that the hybrid approach both saves space and improves efficiency in Section~\ref{sec-expt}.

\stitle{Summary}.
(1) We have shown that the {\em direct} application of distance landmarks as~\cite{PotamiasBCG09} is impractical for large-scale graphs.
(2) We have also proposed a notion of hybrid landmark covers for reducing both space and time cost.
(3) In the following sections, we will explore novel techniques to significantly reduce the space and time cost. These techniques combined with distance landmarks support efficient shortest distance queries on large real-life graphs.}

\section{Using Representatives for Landmarks}
\label{sec-agent}

As illustrated and analyzed in Section~\ref{sec-analysis}, the direct application of distance landmarks is not practical for large graphs.
A straightforward approach is to use {\em representatives}, each of which captures a set of nodes in a graph.
The distance landmarks are for the representatives only, instead of the entire graph, which  reduces both space and time cost.

The task to find a proper form of representatives is, however, {\em nontrivial}. Intuitively, we expect representatives to have the following properties.
(1) A small number of representatives can represent a large number of nodes in a graph;
(2) Shortest distances involved within the set of nodes being represented by the same representative can be answered efficiently; And, moreover,
(3) the representatives and the set of nodes being represented can be computed efficiently.

In this section, we first propose {\em agents} and {\em deterministic routing areas} (\dras) to capture representatives and the set of
nodes being represented, respectively. We then give an analysis of the properties of  \dras and their agents, and show that they are indeed what we want.
 Finally, we present a linear-time algorithm for computing agents and their \dras.
The idea of using agents and \dras is illustrated in Fig.~\ref{fig-angent-landmarks}.

We consider a graph $G(V, E, w)$.

\eat{\begin{figure}[tb!]
\begin{center}
\includegraphics[scale=0.45]{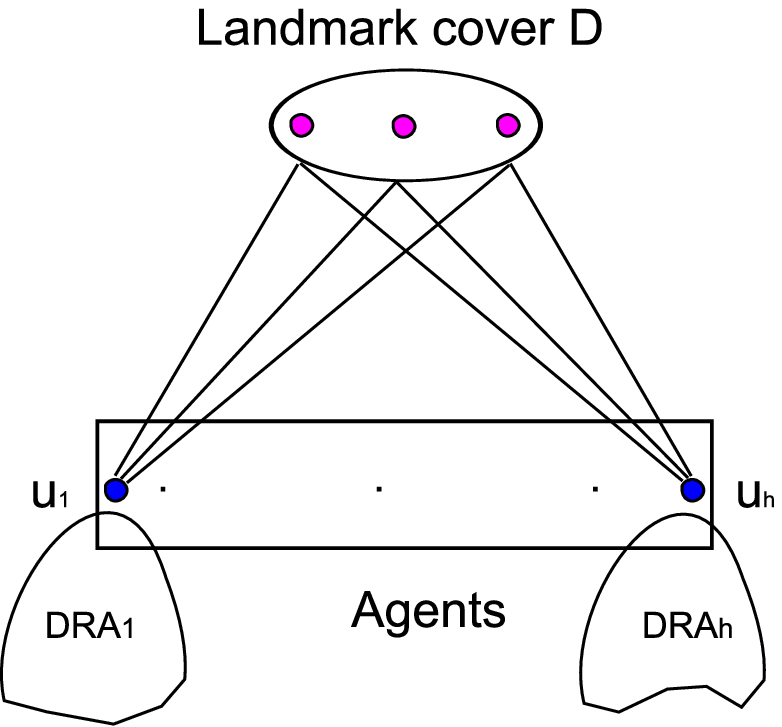}
\end{center}
\vspace{-3ex}
\caption{Using representatives for landmarks}
\label{fig-angent-landmarks} \vspace{-4ex}
\end{figure}
}

\begin{figure}
\centering
\begin{minipage}[b]{.24\textwidth}
  \centering
  \includegraphics[scale=0.45]{./rep-landmarks.eps}
  \caption{Using agents for landmarks}
  \label{fig-angent-landmarks}
\end{minipage}
\begin{minipage}[b]{.24\textwidth}
  \centering
  \includegraphics[scale=0.65]{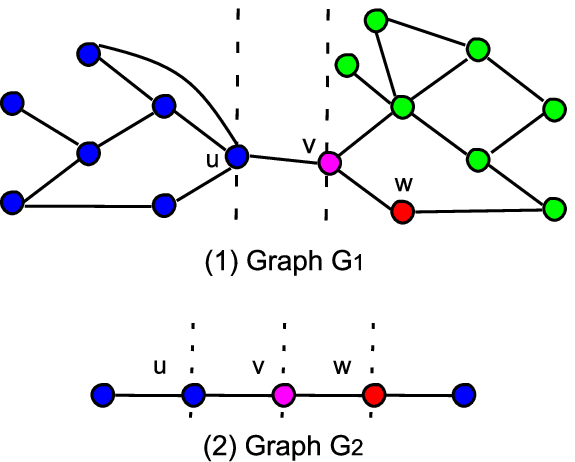}
  \caption{Example agents and \dras}
  \label{fig-agents}
\end{minipage}%
\vspace{-4ex}
\end{figure}

\subsection{Agents and Deterministic Routing Areas}
\label{subsec-agent-def}

We first present agents and their \dras.

\etitle{Agents}. Given a node $u$ in graph $G(V, E)$, we say that $u$ is an {\em agent} of a set of nodes, denoted by $A_{u}$, if and only if:

\sstab(1) node $u\in A_{u}$ is reachable to any node of $A_u$ in $G$,

\sstab(2) all neighbors of any node $v\in A_u\setminus \{u\}$ are in $A_u$,  and

\sstab(3) the size $|A_u|$ of $A_u$ is equal or less than $c\cdot\lfloor\sqrt{|V|}\rfloor$,

\ni where $c$ is a small constant number, such as $2$ or $3$.

Here condition (1) guarantees the connectivity of subgraph $G[A_u]$,  condition (2) implies that not all neighbors of agent $u$ are necessarily in $A_u$;
and condition (3), referred to as {\em size restriction}, limits the size of $A_u$ of agent $u$.

Note that a node $u$ may be an agent of multiple sets of nodes $A^1_u, \ldots, A^k_u$ such that $A^i_u\cap A^j_u$ = $\{u\}$ for any $i\ne j\in[1, k]$.
And we denote as $A^{+}_u$ the union of all the sets of nodes whose agent is $u$ , \ie  $A^{+}_u$ = $A^1_u$ $\cup\ldots\cup$ $A^k_u$.

\etitle{Maximal agents}.  We say that an agent $u$ is {\em maximal} if there exist no other agents $u'$ such that $A^+_{u} \subset A^+_{u'}$.

\etitle{Trivial agents}. We say that a maximal agent $u$ is {\em trivial} if $A^+_u$ contains itself only, \ie $A^+_{u}$ = $\{u\}$.

\etitle{Equivalent agents}. We say that two agents $u$ and $u'$ are {\em equivalent}, denoted by $u\equiv u'$, if $A^+_{u} = A^+_{u'}$.

\etitle{Deterministic routing areas (\dras)}. We refer to the {\em subgraph} $G[A^+_u]$ with nodes $A^+_u$ as a \dra of agent $u$.

Intuitively, \dra $G[A^+_u]$ is a {\em maximal} connected subgraph connecting to the rest of graph $G$ through agent $u$ only.

We next illustrate these notions with an example below.

\vspace{-0.5ex}
\begin{example}
\label{exm-agents} First consider graph $G_1(V_1, E_1)$ in Fig.~\ref{fig-agents}, and let $c\cdot\lfloor\sqrt{|V_1|}\rfloor$ =
$2\cdot\lfloor\sqrt{|16}\rfloor$ = $8$, where $c = 2$ and $|V_1| = 16$.

\sstab(1) Node $u$ is an agent, and its \dra is the subgraph in the left hand side of the vertical line across $u$;

\sstab(2) Node $v$ is an agent, and its \dra is the subgraph in the left hand side of the vertical line across $v$;

\sstab(3) Node $w$ is not an agent since it can not find a \dra with size less or equal than $8$;

\sstab(4)  Node $v$ is a maximal agent, while node $u$ is not a maximal agent since $A^+_u\subset A^+_v$.

\vspace{1ex} We then consider graph $G_2(V_2, E_2)$ in Fig.~\ref{fig-agents}, and let $c\cdot\lfloor\sqrt{|V_2|}\rfloor$ =
$2\cdot\lfloor\sqrt{5|}\rfloor$ = $4$, where $c = 2$ and $|V_2| = 5$.

\sstab(1) Nodes $u, v$ and $w$ are three maximal agents, whose \dras are all the entire graph $G_2$, and, hence,

\sstab(2) $u, v$ and $w$ are three equivalent agents.
 \end{example}
\vspace{-0.5ex}

\vspace{-1ex}
\stitle{Remarks}. (1) As illustrated by the above examples,  a \dra of graph $G(V, E)$ may have a size larger than $c\cdot\lfloor\sqrt{|V|}\rfloor$,
and multiple equivalent agents. (2) Trivial agents can only represent themselves. Hence, we are only interested in non-trivial agents (or simply called agents) in the sequel.

\eat{\begin{figure*}
\centering
\begin{minipage}[b]{.4\textwidth}
  \centering
  \includegraphics[scale=0.65]{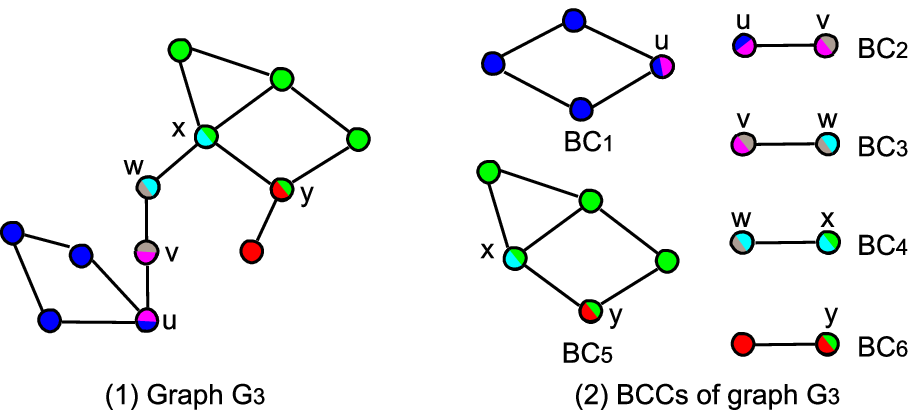}
  \caption{Cut-nodes and bi-connected components}
  \label{fig-cut-nodes}
\end{minipage}
\quad\quad\quad\quad
\begin{minipage}[b]{.4\textwidth}
  \centering
  \includegraphics[scale=0.65]{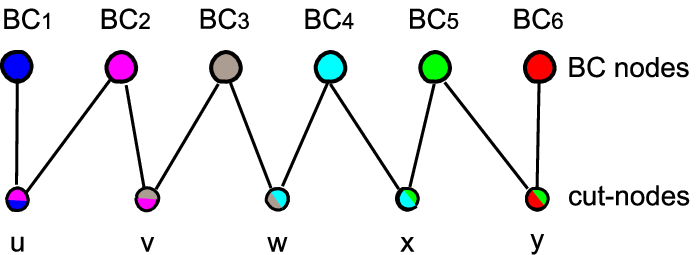}
  \caption{\bcsketch graph $\mathbb{G}_3$ of graph $G_3$}
  \label{fig-sketch-graph}
\end{minipage}
\vspace{-3ex}
\end{figure*}
}

\subsection{Properties of Agents and DRAs}
\label{subsec-agent-properties}

We next give an analysis of agents and \dras, and show that they hold good properties for shortest distance queries.

\begin{prop}
\label{prop-agent-unique-dra} Any agent in a graph has a unique \dra.
\end{prop}

This shows that agents and \dras are well defined notions.

\begin{prop}
\label{prop-agent-cc} Without the size restriction, any node $u$ in graph $G$ is a maximal agent,
and its \dra $G[A^+_u]$ is exactly the connected component (\cc) to which $u$ belongs.
\end{prop}

This justifies the necessity of the size restriction for agents.  Otherwise, \dras are simply \ccs, and are mostly useless.

\begin{prop}
\label{pro-agent-distance} For any two nodes $v, v'$ in the \dra $G[A^+_u]$ of agent $u$ in graph $G$, \\
(1) the shortest distance $\dist(v, v')$ in \dra $G[A^+_u]$ is exactly the one in the entire graph $G$; and\\
(2) it can be computed in linear time in the size of $G$.
\end{prop}

The size restriction guarantees that the shortest distance computation within a \dra can be evaluated efficiently.

\begin{prop}
\label{pro-agent-path} Given a node $v$ in  the \dra $G[A^+_{u}]$ of agent $u$ in graph $G$, and another node $v'$ in $G$, but not in $G[A^+_u]$, the shortest distance
$\dist(v, v')$ = $\dist(v, u)$ $+$ $\dist(u, v')$.
\end{prop}

Propositions~\ref{pro-agent-distance} and~\ref{pro-agent-path} together guarantee that the shortest distances between
 the nodes in the \dras of two distinct agents can be answered correctly and efficiently.

\begin{prop}
\label{prop-agent-cut} Any agent in a \cc $H(V_s, E_s)$ of graph $G(V$, $E)$ with $|V_s|>c\cdot\lfloor\sqrt{|V|}\rfloor$ must be a cut-node of graph $G$.
\end{prop}

This motivates us to identify maximal agents by utilizing the cut-nodes and \bccs, which will be seen immediately.

\begin{prop}
\label{prop-large-bcc} Any node in a bi-connected component (\bc)  with size larger than $c\cdot\lfloor\sqrt{|V|}\rfloor$ of graph $G(V$, $E)$ is a trivial agent.
\end{prop}

As we are interested in non-trivial agents only, those large \bccs could be simply ignored with any side effects.

\begin{theorem}
\label{thm-agent-disjoint} Given any two agents $u$ and $u'$, \\
(1) if $u\in A^+_{u'}$, then $A^+_{u}\subseteq A^+_{u'}$; \\
(2) if $u'\in A^+_{u}$, then $A^+_{u'}\subseteq A^+_{u}$;  and \\
(3) $A^+_{u}\cap A^+_{u'}$ = $\emptyset$, otherwise.
\end{theorem}

\vspace{-1ex}
\begin{cor}
\label{cor-agent-disjoint} Given any two maximal agents $u$ and $u'$, then either $A^+_{u} = A^+_{u'}$ or $A^+_{u}\cap A^+_{u'}$ = $\emptyset$ holds.
\end{cor}

This says when maximal agents are concerned, there exists a unique set of non-overlapping \dras.

\subsection{Computing DRAs and Maximal Agents}
\label{subsec-agent-algorithms}

In this section, we first present a notion of  \bcsketch graphs, based on which we then propose an algorithm for computing \dras and their maximal agents.

The main result here is stated as follows.

\begin{theorem}
\label{thm-compute-dras} Finding all \dras, each associated with one maximal agent, in a graph can be done in linear time.
\end{theorem}

We shall prove this by providing a linear time algorithm that computes \dras and maximal agents.
We first present \bcsketch graphs, a key notion employed by the algorithm.

\etitle{A \bcsketch graph} $\mathbb{G(V, E}, \omega)$ of a graph $G(V, E)$ is a bipartite graph, in which (1) $\mathbb{V}$ = $\mathbb{V}_{c}\cup \mathbb{V}_{bc}$
such that $\mathbb{V}_{c}$ is the set of cut-nodes in $G$, and $\mathbb{V}_{bc}$ is the set of \bccs in $G$;
(2) for each cut-node $v\in \mathbb{V}_{c}$ and each \bc $y_{b}\in \mathbb{V}_{bc}$, there exists an edge $(v, y_b)\in \mathbb{E}$ iff $v$ is a cut-node of \bc $y_b$;
and (3) $\omega$ is a weight function such that for each node $y_b\in \mathbb{V}_{bc}$, $\omega(y_b)$ is the number of nodes of $G$ in \bc $y_b$.

\begin{figure}[tb!]
\begin{center}
\includegraphics[scale=0.7]{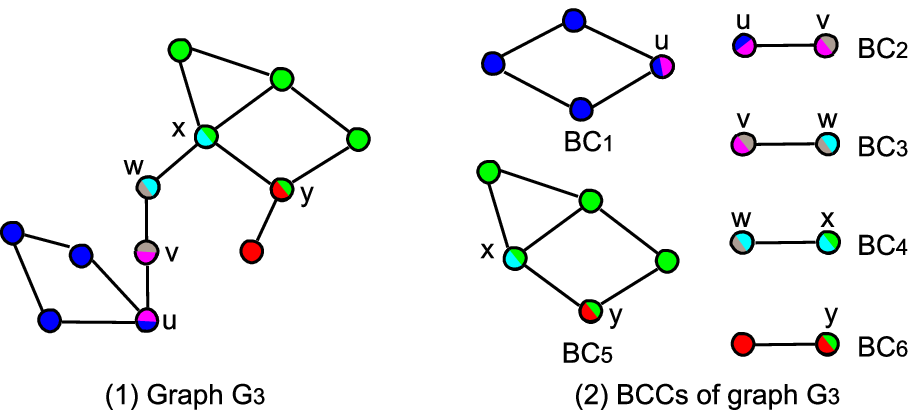}
\end{center}
\vspace{-2ex}
\caption{Cut-nodes and bi-connected components}
  \label{fig-cut-nodes}\vspace{-2ex}
\end{figure}
\begin{figure}[tb!]
\vspace{2ex}
\begin{center}
\includegraphics[scale=0.7]{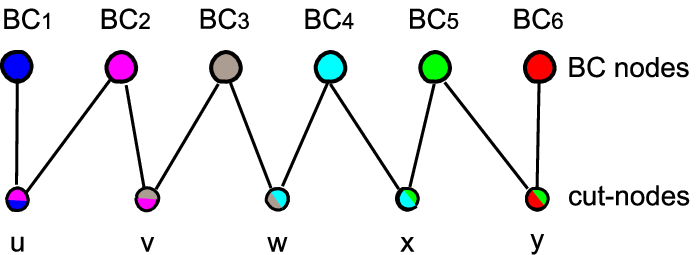}
\end{center}
\vspace{-2ex}
\caption{\bcsketch graph $\mathbb{G}_3$ of graph $G_3$}
\label{fig-sketch-graph}\vspace{-4ex}
\end{figure}

\vspace{-0.5ex}
\begin{example}
\label{exm-sketch-graph} Consider graph $G_3$ in Fig.~\ref{fig-cut-nodes}(1), in which labeled nodes $u, v, w, x, y$ are the cut-nodes of $G_3$,
and the corresponding \bccs of $G_3$ are $BC_1, BC_2, BC_3, BC_4, BC_5$, and $BC_6$, and are shown in Fig.~\ref{fig-cut-nodes}(2).
\vspace{-0.5ex}

The \bcsketch graph $\mathbb{G}_3(\mathbb{V, E}, \omega)$ of graph $G_3$ is shown in Fig.~\ref{fig-sketch-graph},
in which $\omega(BC_1)$ = 4, $\omega(BC_2)$ = $\omega(BC_3)$ = $\omega(BC_4)$ = $\omega(BC_6)$ = 2, and $\omega(BC_5)$ = 5.
\end{example}

One may notice that there are no cycles in the \bcsketch graph $\mathbb{G}_3$. This is not a coincidence, as shown below.

\begin{prop}
\label{pro-sketch-graph} \bcsketch graphs have no cycles, which implies that they are simply trees.
\end{prop}

Proposition~\ref{pro-sketch-graph} indicates that we can employ the good properties of trees for computing \dras and maximal agents.

\begin{figure}[t]
\vspace{1ex}
\begin{center}
{\small
\begin{minipage}{3.36in}
\myhrule \vspace{-2ex}
\mat{0ex}{
\sstab {\sl Input:\/} \= Graph $G(V, E)$ and constant $c$.\\
{\sl Output:\/} The \dras associated with their maximal agents. \\

\sstab\bcc\ \= Find all cut-nodes $\mathbb{V}_c$ and \bc nodes $\mathbb{V}_{bc}$ of $G$; \\
\icc\> Build the \bcsketch graph $\mathbb{G(V, E}, \omega)$ with $\mathbb{V}$ = $\mathbb{V}_c\cup \mathbb{V}_{bc}$;\\
\icc\> Identify and return the \dras and their maximal agents of $G$.
}

\vspace{-2.5ex}
\mat{0ex}{
{\bf Procedure}~$\kw{extractDRAs}$\\
\sstab {\sl Input:\/} \= \bcsketch graph $\mathbb{G(V, E}, \omega)$ of graph $G$ and constant $c$.\\
{\sl Output:\/} The \dras and their maximal agents of $G$.\\
\bcc \hspace{1.8ex}\= \Let $F$ be the set of cut-nodes with leaf neighbors in $\mathbb{G}$;\\
\>  /* {\small note that a leaf node must be a \bc node} */\\
\icc \> \While $F$ is \Not empty \Do\\
\icc \> \hspace{1ex} \= pick a cut-node $v$ from $F$; \ \ \Let $X$ be the neighbors of $v$;\\
\>\>  /* {\small note that there is at most one non-leaf node in $X$} */\\
\icc \>\> \Let $\alpha$ := $\sum_{y'\in X}\omega(y')$ - $|X|$ + 1;\\
\icc \>\> \If $\alpha \le c\cdot\lfloor\sqrt{|V|}\rfloor$ \Then\\
\icc \>\>\hspace{2ex}\=merge all \bc nodes in $X$ and $v$ into one \bc node $y_{n}$;\\
\icc \>\>\> \Let $\omega(y_{n})$ := $\alpha$;\\
\icc \>\>\> \If there is a non-leaf node in $X$ \Then replace it with $y_{n}$;\\
\icc \>\> $F$ := $F\setminus \{v\}$;\\
\icc \> \Let $F'$ be the set of new cut-nodes with leaf neighbors;\\
\icc \> \For each cut-node $v$ in $F'$ \Do \\
\icc \>\>\Let $X'$ be a set of leaf neighbors of $v'$ such that  \\
\icc \>\> for each $y'\in X'$, $\omega(y')$ $\le$ $c\cdot\lfloor\sqrt{|V|}\rfloor$;\\
\icc \>\> mark $X'$ as the \dra $A^+_{v'}$ of agent $v'$;\\
\icc \>\Return all \dras with their maximal agents.}
\vspace{-2.5ex} \myhrule
\end{minipage}
}
\end{center}
\vspace{-2ex}
\caption{Computing \dras and maximal agents}
\label{alg-compute-agents}
\vspace{-4ex}
\end{figure}

We are now ready to present algorithm \compDRAs shown in Fig.~\ref{alg-compute-agents}.
It takes as input graph $G$ and constant $c$, and outputs the \dras of $G$, each associated with a maximal agent.

\etitle{(1) Finding cut-nodes and \bccs}.
The algorithm starts with computing all cut-nodes and bi-connected components (line 1), by using the linear-time algorithm developed by
John Hopcroft and Robert Tarjan~\cite{CormenLRS01,HopcroftT73-cut-nodes}.

\etitle{(2) Constructing \bcsketch graphs}.
After all the cut-nodes and \bccs are identified, the \bcsketch graph $\mathbb{G(V, E}, \omega)$ can be easily built (line 2).
To see this can be done in linear time, the key observation is that the number $|\mathbb{E}|$ of edges in $\mathbb{G}$ is
exactly $|\mathbb{V}| - 1$ since $\mathbb{G}$ is a tree.

\etitle{(3) Identifying \dras and their maximal agents}.
Finally, the algorithm identifies and returns the \dras and their maximal agents (line 3), using
Procedure~\kw{extractDRAs} in Fig.~\ref{alg-compute-agents}.

\etitle{Procedure~\kw{extractDRAs}} takes as input the \bcsketch graph $\mathbb{G}$ of graph $G$ and constant $c$,
and outputs the \dras and their maximal agents, by repeatedly merging \bccs with size less than $c\cdot\lfloor\sqrt{|V|}\rfloor$.
More specifically, the procedure starts with the set $F$ of cut-nodes with leaf neighbors (line 1).
It then recursively merges the neighboring \bc nodes of cut-nodes to generate new \bc nodes (lines 2-9).
For a node $v\in F$ with neighbors $X$, if $\sum_{y'\in X}\omega(y')$ - $|X|$ + $1$ $\le$ $c\cdot\lfloor\sqrt{|V|}\rfloor$,
they can be merged into a new \bc node (lines 3-8). Intuitively, this says cut-node $v$ is not a maximal agent, and it is combined into the \dras of maximal agents.
A key observation here is that there is at most one non-leaf node in $X$.
If there is such a non-leaf neighbor, then it is replaced by the new \bc node $y_n$ (line 8), by which the merging processing is made possible.
Once a cut-node is considered, it is never considered again (line 9).
After no merging can be made, we have found all maximal agents, \ie all the cut-nodes in the updated \bcsketch graph.
We then identify \dras for these maximal agents (lines 10-14). For any leaf neighbor $y'$ of a cut-node $v'$,
if $\omega(y')$ $\le$ $c\cdot\lfloor\sqrt{|V|}\rfloor$, then $y'$ is an $A_{v'}$ of agent $v'$.
All these together constitute the $A^+_{v'}$ of agent $v'$ (lines 12-14). Finally, all \dras with their maximal agents are returned (line 15).

We now explain the algorithm with an example as follows.

\vspace{-0.5ex}
\begin{example}
\label{exm-compute-agents} Consider graph $G_3$ in Fig.~\ref{fig-cut-nodes}(1) again. Here we let $c = 2$, and $c\cdot\lfloor\sqrt{|V|}\rfloor$ = $6$.
Firstly, cut-nodes and \bccs are computed as shown in Fig.~\ref{fig-cut-nodes}(2).
Secondly, the \bcsketch graph $\mathbb{G}_3$ of $G_3$ is constructed as shown in Fig.~\ref{fig-sketch-graph}.
After the merging step stops, the updated \bcsketch graph consists of three \bc nodes:
$BC'_1$ = $\{BC_1, BC_2, BC_3\}$, $BC_4$, $BC'_2$ = $\{BC_5, BC_6\}$ and two cut-nodes: $w$ and $x$.
Finally, the \dras and their maximal agents are identified: agent $w$ with \dra $BC'_1$ and agent $x$ with  \dra $BC'_2$.
\end{example}
\vspace{-0.5ex}

\vspace{-1ex}
\stitle{Correctness \& Complexity}. The correctness of algorithm \compDRAs can be readily verified based on the analyses in Section~\ref{subsec-agent-properties}.
To show that algorithm $\compDRAs$ runs in linear time,
it suffices to show that procedure $\kw{extractDRAs}$ can be done in linear time.
It is easy to see that each node in the \bcsketch graph is visited at most twice in procedure $\kw{extractDRAs}$, and hence the procedure runs in linear time.

This completes the proof of Theorem~\ref{thm-compute-dras}.

\stitle{Summary}. (1) We have proposed a notion of agents and \dras aiming at reducing the size of graphs such that
landmarks are only for agents, instead of the entire graph.
(2) We have given a theoretical analysis of agents and \dras, based on which we have developed a linear time algorithm for computing \dras and their maximal agents.
(3) As shown in our experimental study, on average about 1/3 nodes of a graph are captured by non-trivial agents and their \dras.

\section{Introducing Graph Partitions for Landmarks}
\label{sec-decomposition}

Web graphs contain a large strongly connected components~\cite{BroderKMRRSTW00}, and, similarly, there is usually a large \bc in real-life graphs such as the collaboration and social networks~\cite{LeskovecLDM08,corr-abs-MF}.
As pointed out in Section~\ref{sec-agent}, for the \bccs in a graph $G(V, E)$ with  a size larger than $\lfloor\sqrt{|V|}\rfloor$, each node in those \bccs is a trivial agent that can only represent itself. This motivates us to introduce the graph partitioning techniques for distance landmarks,
based on which we use a small set of nodes, instead of a single agent node, to represent a large set of nodes.

In this section, we first introduce graph partitions.
We then propose a notion of \super graphs which combine graph partitions with hybrid landmark covers.
We finally present the bounded graph partition problem and its solution.

We consider a graph $G(V, E)$.

\subsection{Graph Partitions and Super Graphs}
\label{subsec-supergraphs}

We first introduce graph partitions and \super graphs.

\etitle{Graph partitions}. We say that $(V_1, \ldots, V_k)$ is a {\em partition} of graph $G(V, E)$ if and only if (1) $\bigcup_{i=1}^k V_i$ = $V$, and (2) for any $i\ne j\in[1,k]$, $V_i\cap V_j = \emptyset$, in which we refer to a $V_i$ ($i\in[1, k]$) as a {\em fragment} of the partition.

We also say  that node $u$ in $V_i$ ($1\le i\le k$) is a {\em boundary} node if there exists an edge $(u, v)$ in $G$ from nodes $u$ to $v$ such that $v\in V_j$ and $j\ne i$ ($1\le j\le k$).

\etitle{\super graphs}. We next introduce \super graphs that combine graph partitions with hybrid landmark covers.

Consider a partition  $(V_1, \ldots, V_k)$ of graph $G$. For each fragment $V_i$ ($i\in[1, k]$), let (1) $B_i$ be the set of boundary nodes of $V_i$, and (2) $\tilde{D_i} = (D_i, E_{D_i}^-)$ be a hybrid landmark cover for the set $B_i$ of {\em boundary} nodes of $V_i$.

The \super graph of graph partition  $(V_1, \ldots, V_k)$ is a weighted undirected graph $\G(\V, \E, \Upsilon)$ such that:

\sstab(1) $\V$ = $B_1\cup\ldots\cup B_k\cup D_1\cup\ldots\cup D_k$, \ie the union of all boundary nodes and distance landmarks on each fragment;

\sstab(2) $E$ = $E_B\cup E_{\tilde{D_1}}\cup\ldots\cup E_{\tilde{D_k}}$, where $E_B \subseteq E$ is the set of edges with both endpoints belonging to $B_1\cup\ldots\cup B_k$, and for each $i\in[1, k]$, $E_{\tilde{D_i}}$ is the set of edges enforced by the hybrid landmark cover $\tilde{D_i}$; and

\sstab(3) For each edge $(u, v) \in E_B$, $\Upsilon(u,v)$ is exactly equal to the edge weight $w(u, v)$ in graph $G$, and  for each edge $(u, v) \in E_{\tilde{D_i}}$ ($i\in[1, k]$), $\Upsilon(u,v)$ is the local shortest distance between $u$ and $v$ in the fragment $V_i$ only.

That is, a \super graph $\G$ of graph $G(V, E)$ only consists of the landmarks and  boundary nodes. Hence, the size of $\G$ is typically much smaller than graph $G$. Intuitively, \super graphs  use a small set of nodes in a fragment, \ie the boundary nodes and distance landmarks, to represent a large number of nodes, \ie all the nodes in the fragment.

\subsection{Bounded Graph Decompositions}
\label{subsec-boundedGD}


As the landmarks are for the boundary nodes, the number of boundary nodes has a key impact on the size of \super graphs.
In addition, the size of a fragment should be bounded in order to efficiently compute its hybrid landmark cover.

This motivates us to study the following problem.

\etitle{The bounded graph partitioning problem} is to find a partition $(V_1, \ldots, V_k)$ of graph $G(V, E)$ , denoted by \gdp, such that (1) $|V_i|\le \Gamma$ for each fragment $V_i$ ($i\in[1, k]$), and (2) $|B|\le \epsilon\cdot|V|$,
where $\Gamma\le |V|$ is a positive integer, $\epsilon\in[0.0,1.0]$ is a rational number, and $|B|$ is the total number of boundary nodes.

The problem is, however, nontrivial, as expected.

\begin{prop}
\label{pro-graph-decomposition} The \gdp problem is \NP-complete.
\end{prop}

Traditional graph partitioning is to find a partition $(V_1$, $\ldots$, $V_k)$  of a graph such that (1) the $k$ fragments have a roughly equal number of nodes,
and (2) the number of edges connecting nodes in different fragments is minimized.
The problem has been extensively studied since 1970's~\cite{kl70,Karypis98,YangYZK12},
and has been used in various applications, \eg circuit placement, parallel computing and scientific simulation~\cite{YangYZK12}.

Large-scale graph partitioning tools are available such as the best-known \metis~\cite{Karypis98}.
Hence, this study is not to propose a new graph partitioning algorithm. Instead, it builds relationships between the \gdp problem and
the traditional graph partitioning problem, and makes use of existing approaches for solving the \gdp problem.

\etitle{Key observations}. For any partition $(V_1$, $\ldots$, $V_k)$,  the set $B$ of boundary nodes with edges across different fragments and the set $E_B$ of
all edges connecting nodes in different fragments satisfy: $|B| \le 2|E_B|$.

This is, minimizing $|E_B|$ essentially reduces the upper bound of $|B|$. Moreover, those edges in $E_B$ are part of the \super graph.
Hence, minimizing $|E_B|$ also reduces the size of the \super graph. This observation inspires us to adopt
existing approaches, \eg\ \metis~\cite{Karypis98}, to partition graphs and generate \super graphs.
As will be seen in in our experiments, smaller \super graphs help answer shortest distance queries.

\stitle{Summary}. (1) We have introduced a notion of \super graphs that combine graph partitions with distance landmarks.
(2) We have proposed the \gdp problem, and shown it is \NP-complete.
(3) We have also built connections between the \gdp problem and the traditional graph partitioning problem,
which makes it possible to use the existing approaches, \eg\  \metis~\cite{Karypis98}, to solve our problem.
As will be seen in our experiments, \metis works well for the \gdp problem, and the produced \super graphs are typical small,
which only have 2--4\% nodes and 10--15\% edges compared with the original graphs.

\section{A Unified Framework for Answering Shortest Distance Queries}
\label{sec-query}

In this section, we propose a unified framework, referred to as \disland, for fast shortest distance query answering, which consists of two modules: {\em preprocessing } and {\em query answering}. We combine distance landmarks with agents and graph partitions (\super graphs), and seamlessly integrate existing speed-up techniques \cite{GeisbergerSSD08, MohringSSWW05} into the framework.

Consider a graph $G(V, E)$ with non-negative edge weights.

\subsection{Preprocessing for Query Answering}
\label{subsec-preprocessing}

We first present the preprocessing module.

Given graph $G(V, E)$, the module seamlessly combines agents and graph partitions with hybrid landmark covers, and it produces (a) maximal agents along with their \dras, (b) graph partitions, and (c) a \super graph $\G(\V, \E)$.

More specifically, given graph $G(V, E)$, the module executes the following processes:

\sstab (1) It first computes the \dras and their maximal agents, using algorithm $\compDRAs$ proposed in Section~\ref{subsec-agent-algorithms}.

\sstab (2) For each \dra with a non-trivial maximal agent $u$, it further (a) computes all the shortest distances $\dist(u, v)$ for all nodes $v$ in its \dra, and
(b) adds an edge $(u, v)$ with weight $\dist(u, v)$ for each node $v$ in the \dra.

\sstab (3) It then generates a {\em shrink graph}, the subgraph $G[A]$ of $G$ in which $A$ is the set of agent nodes, including both trivial and non-trivial agents.
For each \dra with a maximal agent $u$, only $u$ is kept in $G[A]$.

\sstab (4) It next calls \metis~\cite{Karypis98} to produce a graph partition $(V_1, \ldots, V_k)$ for the shrink graph $G[A]$ such that
    for each $i\in[1, k]$, $|V_i|$ is roughly equal to $c\cdot\lfloor\sqrt{|V|}\rfloor$. Here $c$ is a small constant number, such as $2$ or $3$.

\sstab (5) For each fragment $V_i$ ($i\in[1, k]$), it computes a (local) hybrid landmark cover $\tilde{D_i}$ for the {\em boundary nodes} of $V_i$ only, by calling the \scover based algorithm (Section~\ref{subsec-dislandmarkdef}, \cite{PotamiasBCG09}). Note that here we did not use the \vcover based algorithm, which was proposed for estimating of the size of landmark covers only.

\sstab (6) Finally, it builds a \super graph $\G(\V, \E, \Upsilon)$ of graph $G$.

\begin{figure}[tb!]
\begin{center}
\includegraphics[scale=0.8]{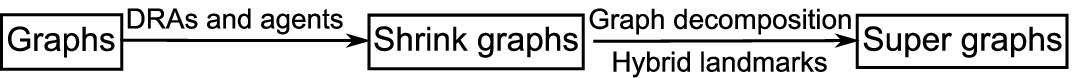}
\end{center}
\vspace{-2ex}
\caption{The preprocessing module}
\label{fig-preprocess} \vspace{-4ex}
\end{figure}

The entire process is illustrated in Fig.~\ref{fig-preprocess}.


\subsection{A Bi-level Query Answering Approach}
\label{subsec-answering}
We next present the query answering  module.

Given a source node $s$ and a target node $t$, this module finds the shortest distance from $s$ to $t$, by making use of the auxiliary structures produced by the preprocessing module.

\eat{
The $(s, t)$ augmented \super graph is simply the \super graph $\mathbb{S}(\mathbb{V, E})$ with augmented edges in $G[V_s]$ and $G[V_t]$.

\begin{example}
\label{exm-super-graphs}
\end{example}
}

More specifically, given nodes $s$ and $t$, the query answering  module executes the following processes:

\sstab(1) When nodes $s$ and $t$ belong to the same \dra $G[A^+_u]$ with agent $u$ such that $A^+_u$ = $A^1_u\cup\ldots A^h_u$.

If $s$ and $t$ further fall into the same $A^i_u$, then it invokes Dijkstra's algorithm on the subgraph $G[A^i_u]$. Otherwise, it simply returns $w(s, u)$ + $w(u, t)$ in constant time.

\sstab(2)  When $s$ and $t$ belong to two \dras $G[A^+_{u_s}]$ and $G[A^+_{u_t}]$ with agents $u_s$ and $u_t$, respectively.
As $\dist(s, t)$ = $\dist(s, u_s)$ + $\dist(u_s, u_t)$ + $\dist(u_t, t)$, in which
$\dist(s, u_s)$ and $\dist(u_t, t)$ are already known, we only need to compute $\dist(u_s, u_t)$.

Let $V_s$ and $V_t$ be the fragments to which agents $u_s$ and $u_t$ belong, respectively.
As observed in \cite{ChanL07}, fragments $V_s$ and $V_t$ and the \super graph together suffice to answer exact shortest distance queries.
Hence, the algorithm invokes the Dijkstra's algorithm
on the union of subgraphs $G[V_s]$, $G[V_t]$ and the \super graph $\G(\V, \E, \Upsilon)$ to compute $\dist(u_s, u_t)$.

Following the analysis above, we have the following.

\begin{prop}
\label{pro-query-correctness} Framework \disland correctly answers shortest distance queries.
\end{prop}

\subsection{Optimization Techniques}
\label{subsec-opt}

There exist quite a few speed-up techniques for shortest distance computations~\cite{WagnerW07,WuXDCZZ12}.
 \disland is very flexible such that most of these techniques, if not all, can be seamlessly incorporated to further speed-up  shortest distance query answering. In this study we have adopted bidirectional search \cite{LubyR89}, contraction hierarchies (\ch) \cite{GeisbergerSSD08}, and Arc-Flags (\arcflag) \cite{MohringSSWW05} due to their effectiveness and generality.

We first introduce the three optimization techniques.

\vspace{-0.5ex}
\etitle{(1) Bidirectional search} (\bisearch, \cite{LubyR89}) simultaneously performs two searches: forward and backward, starting at the source and target nodes, respectively \cite{LubyR89,WagnerW07}. It invokes two instances of the Dijkstra's algorithm simultaneously, and  has the same time complexity as the (single directional) Dijkstra's algorithm. However,  \bisearch is usually more efficient than the Dijkstra's algorithm in practice.

\vspace{-0.5ex}
\etitle{(2) Contraction hierarchies} (\ch, \cite{GeisbergerSSD08})  first imposes a total order $\mathcal{O}$ on the nodes of a graph, in ascending order of their relative `importance', and then constructs a {\em hierarchy} by contracting all the nodes in this order.
A node $v$ is contracted by removing it from the graph such that shortest paths in the remaining graph are preserved, achieved by replacing paths of the form $u/v/w$ by a {\em shortcut} edge $(u, w)$. Note that the shortcut $(u, w)$ is {\em only} required if $u/v/w$ is the {\em only} shortest path from $u$ to $w$. After all the nodes are contracted, all the shortcuts are appended into the graph.

\ch uses \bisearch with minor revisions for query answering. Give two nodes $u$ and $w$ with $\mathcal{O}(u)<\mathcal{O}(w)$, \ch only visits two kinds of paths $u/\cdots/v_i/\cdots/w$ in the process: (a) $\mathcal{O}(u)$ $<$ $\cdots$ $<$ $\mathcal{O}(w)$ or
(b) there is a unique $v_i$ with $\mathcal{O}(u)$ $<$ $\cdots$ $<$ $\mathcal{O}(v_i)$ and $\mathcal{O}(v_i)$ > $\cdots>\mathcal{O}(w)$. In this way, \ch avoids visiting the nodes with an order lower than $u$ and $w$ in the forward and backward searches, respectively, which makes it much more efficient than \bisearch alone in practice.

\etitle{(3) Arc-Flags} (\arcflag, \cite{MohringSSWW05}) is a partition-based edge labeling approach, and it divides a graph $G(V, E)$ into partitions $(V_1, \ldots, V_k)$ and gathers information for each edge $e\in E$ and for each fragment $V_i$ ($i\in[1, k]$) on whether the edge $e$ lies on a shortest path into the fragment $V_i$.
To do this, each edge $e$ is associated with a flag vector $f_e$ with $k$ bits (the number of fragments) such that the vector $f_e$ contains
a flag 1 or 0 for $V_i$ indicating whether or not $e$ is useful for a shortest path query to nodes in $V_i$. It is easy to verify that \arcflag incurs $k|E|$ bits of extra space.

\vspace{0.5ex}
We next show how to seamlessly incorporate these three optimization techniques into our framework \disland.

\sstab (1) The shrink graph $G[A]$ of graph $G$ is appended with shortcuts, by using the \ch approach.

\sstab (2) We build a hybrid landmark cover for each fragment, by incorporating the \ch searching process.

We only consider the shortest paths $\rho$ = $u/\cdots/v_i/\cdots/w$ such that (a) $\mathcal{O}(u)<\cdots<\mathcal{O}(w)$, in which case $\rho$ is called {\em order rising}, or (b) $\mathcal{O}(u)$ < $\cdots<\mathcal{O}(v_i)$ and $\mathcal{O}(v_i)$ > $\cdots>\mathcal{O}(w)$, in which case $\rho$ is called {\em order turning}.  When computing landmarks for a fragment, we cover a node pair $(u,v)$ only if (1) there exists an order rising or turning path between $u$ and $v$, and (2) their (local) shortest distance in the fragment is equal to their (global) shortest distance in the entire shrink graph. Moreover, (a) for these node pairs $(u, w)$ connected by order turning paths, we select the nodes with {\em highest order} as landmarks; and (b) for these remaining node pairs $(u, w)$ connected by order rising paths, we use the cost model to greedily select landmarks or build direct edges, following the hybrid landmark approach.
As the searching space is reduced, this both improves the efficiency of computing hybrid landmark covers, and, of course, the query answering. Moreover, we adopt the query answering approach for \ch \cite{GeisbergerSSD08}, instead of the bidirectional Dijkstra's algorithm, in the query answering module of \disland.

\eat{Written by Kaiyu
According to the CH searching rules, we define that a path $P = <v_1, v_2, \dots, v_k>$ is a rising path if $order(v_1) < order(v_2) < \dots < order(v_k)$, and a turning path if there exist a $i \in (1,k)$ such that $order(v_1) < \dots < order(v_i) > \dots > order(v_k)$. In each fragment, we only cover a node pair $(u,v)$ only if (1) there exists a rising path or a turning path between the two nodes and (2) the length of the path is equal to the distance between the two nodes. For those node pairs which are connected by a turning path, we select the node with the maximum order in the turning path to be a landmark. For those node pairs which are connected by a rising path, we will use a cost model to select landmarks greedily in our hybrid landmarks method. We add shortcuts to connect the uncovered node pairs when we can't select a node with an acceptable cost any more. When using the pure landmark method, we do the same thing as we do in hybrid landmark method to the node pairs which are connected by a turning path. We think each node in the rising path can cover the node and transform the landmark selecting problem into a set cover problem and use a greedy strategy to solve it.

Let $t$ be the number of edges enforced by the hybrid landmark cover in all fragment. We need to store the destination and the weight of this edge. So we need $4\cdot 2\cdot t$ byte space to store all these enforced edges. From table 8 we can see that hybrid landmarks covers cost less space than landmarks covers
}

\sstab (3) We compute edge labeling, by using the \arcflag approach. To do this, we further call \metis to do a second level partition of the \super graph, where each fragment is treated as a single node, and the edge weight between fragments are the number of edges connecting them. When building Arc-Flags, we again incorporate \ch, by considering order rising or turning shortest paths only, to speed-up the processing.


\stitle{Extra space analysis}. This module produces two kinds of auxiliary structures: the non-trivial maximal agents along with their \dras
and the \super graph $\G(\V, \E, \Upsilon)$.

\sstab(1) Let $U = \{u_1, \ldots, u_h\}$ be the set of non-trivial maximal agents identified.
The extra space of $U$ and their \dras is the extra edges from those agents to the set of nodes in their \dras, which is exactly equal to $\sum_{i=1}^h|A^+_{u_i}|$ - $h$.

\sstab(2) Each fragment in the partition $(V_1, \ldots, V_k)$ roughly has the same size of $c\cdot\lfloor\sqrt{|V|}\rfloor$. We set $c$ = $2$ or $3$ in practice.
Hence, the number of fragments is less than $\lfloor\sqrt{|V|}\rfloor$.

For each fragment $V_i$ ($i\in[1, k]$), let $E_{\tilde{D_i}}$  be the set of edges enforced by the hybrid landmark cover $\tilde{D_i}$ for the {\em boundary} nodes of $V_i$. Hence, the number of extra edges in the \super graph $\G$ is bounded by $\sum_{i=1}^k|E_{\tilde{D_i}}|$.

\sstab(3) The remaining extra space is incurred by the shortcuts added by \ch and the Arc-Flags added by \arcflag.

As will be shown in our experiments, all these auxiliary structures only incur a small space cost, and the entire preprocessing can be finished in a reasonably fast way.

\eat{

Putting these together, the total extra space is bounded by $O(\sum_{i=1}^h|A^+_{u_i}|$ - $h$ +  $\sum_{i=1}^k|E_{\tilde{D_i}}|)$.

\textbf{Todo:} add how to combine with \ch.

\stitle{Summary}.

(1) first do \ch, then do \disland

(2) first do \disland, then do \ch

(3) multi-level super graphs

}



\section{Experimental Study}
\label{sec-expt}

We next present an extensive experimental study of the \disland framework for shortest distance query answering.
Using real-life road networks, we conducted five
sets of experiments to evaluate:
(1) the impacts of agents, graph partitions, and hybrid landmark covers;
(2) the preprocessing time and space overhead of bidirectional Dijkstra~\cite{LubyR89}, \ch~\cite{GeisbergerSSD08}, \arcflag \cite{MohringSSWW05}, their counterparts using agents (Agent + Dijkstra, Agent + \ch, Agent + \arcflag), and \disland; and (3) the performance of all these approaches.

\subsection{Experimental Settings}
We first introduce the settings of our experimental study.

\begin{table}[t!]
\label{tab-datasets}
\caption{Real-world graphs}
\begin{center}
\begin{scriptsize}
\vspace{-2ex}
\begin{tabular}{|c|c|r|r|}
\hline
  Name                            &  Regions               & \# of Nodes  &  \# of Edges \\
\hline\hline
CO      &  Colorado              & 435,666      & 1,042,400  \\ \hline
FL      &  Florida               & 1,070,376    & 2,687,902  \\ \hline
CA      &  California \& Nevada   & 1,890,815   & 4,630,444  \\ \hline
E-US    &  Eastern US            & 3,598,623    & 8,708,058 \\ \hline
W-US    &  Western US            & 6,262,104    & 1,5119,284  \\ \hline
C-US    &  Central US            & 14,081,816   & 33,866,826 \\ \hline
US      &  Entire US             & 23,947,347   & 57,708,624  \\ \hline
\end{tabular}
\vspace{-4ex}
\end{scriptsize}
\end{center}
\vspace{-4ex}
\end{table}

\etitle{Real-life graphs}. We chose seven datasets of various sizes from the Ninth DIMACS
Implementation Challenge~\cite{dimacs-datasets}, shown in Table~2.
Each dataset is an undirected graph that represents a part of the road network in the United States (US), where
each edge weight is the distance (integers) required to travel between the two endpoints of the edge.

\etitle{Distance queries}. We adopted the query generator in~\cite{WuXDCZZ12}. Our distance queries were generated as following.
On each road network, we generated eight sets $Q_1$, $Q_2$, $\dots$ , $Q_{8}$ of
queries. (1) We first imposed a $256 \times 256$ grid on the
road network and computed the side length $\ell$ of each grid cell.
(2) We then randomly chose ten thousand node pairs from
the road network to compose $Q_i (i \in [1, 8])$, such that the grid
distance of all node pairs in $Q_i$ is in $[2^{i-1}\cdot\ell, 2^i\cdot\ell)$. Note
that the grid distance of two nodes $u, v$ in a query set is the distance of the cells into which $u$ and $v$ fall, respectively.
Moreover, the grid distance of any node pair in $Q_i$ is
larger than the grid distance of all node pairs in $Q_{i-1}$.
For each query set $Q_i$ $(i \in [1, 8])$, we report the average running time of over all the ten thousand queries in the set.

\etitle{Algorithms}. We adopted the latest version $5.0.2$ of \metis \cite{metis}, implemented with ANSI C. We also re-implemented the original \ch~\cite{ch-algorithm} from its inventors of using Microsoft Visual C++.  Bidirectional Dijkstra, \arcflag and their counterparts using agents were also written in Microsoft Visual C++. All these algorithms used common data structures and procedures, borrowed from \ch~\cite{ch-algorithm}, for similar tasks.

All experiments were run on a PC with an Intel Core i5-2400 CPU@3.10GHz and 16GB of memory. Each test
was repeated over 5 times, and the average is reported here. We compare algorithms running on general commercial PCs with a 16GB memory limitation, and hence, algorithms using larger memory, \eg~\cite{AbrahamDGW11}, are not in our consideration.

\subsection{Experimental Results}
We next present our findings. In all experiments, we tested the datasets in Table~2, and fixed the constant $c = 2$ when computing agents and graph partitions on graphs $G(V, E)$.

\begin{table}[t!]
\label{tab-exp1-agents-dras}
\caption{Effectiveness of agents and \dras}
\begin{center}
\begin{scriptsize}
\vspace{-2ex}
\begin{tabular}{|c|r|r|r|}
\hline
  Graphs   &  Agents  (\#, \%) &  Nodes (\#, \%) in \dras  & $\timec$ (s) \\
\hline\hline
CO      &  (56,277, 12.9\%)       & (156,329, 35.9\%)     &  1.1  \\ \hline
FL      &  (140,379, 13.1\%)      & (378,937, 35.4\%)     &  3.7 \\ \hline
CA      &  (273,191, 14.4\%)      & (623,811, 33.0\%)     &  11.3 \\ \hline
E-US    &  (546,481, 15.2\%)      & (1,228,876, 34.1\%)   &  34.3  \\ \hline
W-US    &  (869,904, 13.9\%)       & (2,116,339, 33.8\%)   &  100.4 \\ \hline
C-US    &  (2,034,358, 14.4\%)    & (4,583,413, 32.5\%)   &  402.4 \\ \hline
US      &  (3,452,222, 14.4\%)    & (7,927,453, 33.1\%)   &  1153.7 \\ \hline
\end{tabular}
\vspace{-4ex}
\end{scriptsize}
\end{center}
\vspace{-4ex}
\end{table}

\etitle{Exp-1: Impacts of agents}. In the first set of experiments, we evaluated (1) the number of non-trivial agents, (2) the number and percentage of the nodes represented by the agents (excluding the agents themselves from \dras), and (3) the efficiency of our algorithm $\compDRAs$ for computing agents and their \dras.
The results are reported in Table~3.

There are around $1/7$ nodes are non-trivial agents, and about $1/3$ nodes are captured by agents in these graphs, which means basically the shrink graph is only about $2/3$ of the input graph. Moreover, although the size restriction is $\le 2\cdot\lfloor\sqrt{|V|}\rfloor$, \dras are typically small in these graphs, and each agent represents 2 or 3 other nodes on average. Algorithm $\compDRAs$ also scales well, and it can be done in less than half an hour for the largest graph in the preprocessing.

As will be seen in the following experiments, this makes agents a light-weight optimization techniques, which benefits most, if not all, existing shortest distance algorithms.


\begin{table}[t!]
\vspace{4ex}
\label{tab-exp2-partitions-c2}
\caption{Effectiveness of graph partitions}
\begin{center}
\begin{scriptsize}
\vspace{-2ex}
\begin{tabular}{|c|r|r|r|r|}
\hline
    Shrink       &    fragments            &  avg \# of    & avg (\#, \%) of &\\
    graphs &    \multicolumn{1}{|c|}{ (\#) }    &  \multicolumn{1}{|c|}{nodes}    & boundary nodes & $\timec$ (s)\\

\hline\hline
CO      &  220    & 1,269.7   & (76.1, 5.99\%)  &  1.1   \\ \hline
FL      &  340    & 2,033.6   & (92.5, 4.55\%)  &  3.1   \\ \hline
CA      &  470    & 2,695.8   & (114.9, 4.26\%) &  6.6   \\ \hline
E-US    &  630    & 3,761.5   & (156.4, 4.16\%) &  13.8  \\ \hline
W-US    &  840    & 4,935.4   & (151.9, 3.08\%) &  26.2  \\ \hline
C-US    &  1,280  & 7,420.6   & (241.4, 3.25\%) &  85.5  \\ \hline
US      &  1,650  & 9,709.0   & (260.2, 2.68\%) &  126.7 \\ \hline
\end{tabular}
\vspace{-4ex}
\end{scriptsize}
\end{center}
\vspace{-3.5ex}
\end{table}

\etitle{Exp-2: Impacts of graph partitions}.
In the second set of experiments, we justified that the \gdp problem could be solved well by \metis, originally for traditional graph partitioning problems. Using the shrink graphs generated at Exp-1, we evaluated the effectiveness and  efficiency of \metis. To ensure the query efficiency of \disland, each fragment has at most $c \cdot \lfloor\sqrt{|V|}\rfloor$ number of nodes. We used the multilevel bisection method of \metis with the balance factor fixed to 1.003. The results are reported in Table~4.

The results tell us that there are only about (up to) 6\% of nodes are boundary nodes, and the largest graph can be finished in 127 seconds. This clearly justified our analysis and choice to attack the \gdp problem by using existing approaches to traditional graph partitioning problems.

\eat{
We compared the partition quality of Metis with $c = 2, 3, 4$. In order to produce  a partition $(V_1, \dots, V_k)$ for the shrink graph and for each $i \in [1,k]$, $|V_i|$ is roughly equal to $c\cdot \lfloor \sqrt{|V|}\rfloor $, we simply set the number of fragments as follows.need to partition each graph into at least $\lceil \frac{|A|}{c \cdot \sqrt{|V|}}\rceil$ fragments. To insure each fragment has less than $c\cdot \lfloor \sqrt{|V|}\rfloor $ nodes, we set the number of fragments as followed. If $\lceil \frac{|A|}{c \cdot \sqrt{|V|}}\rceil$ mod $10$ $>5$, then we partition the graph into $\lceil \frac{|A|}{c \cdot \sqrt{|V|}\cdot 10} \rceil \cdot 10 + 20$ fragments, otherwise, we partition the graph into $\lceil \frac{|A|}{c \cdot \sqrt{|V|}\cdot 10} \rceil \cdot 10 + 10$.

We used the multilevel recursive bisectioning method provided by Metis. We used the sorted heavy-edge matching to be the matching scheme during coarsening and greedy strategy to grow a bisection during initial partitioning. We also used FM-based cut refinement for refinement. The ufactor was set to be 3 and the number of balancing constraints was set to be 1. All other parameters were set as the default value.
}

\begin{table}[t!]
\label{tab-exp2-landmarks}
\caption{Effectiveness of hybrid landmark covers}
\begin{center}
\begin{scriptsize}
\vspace{-2ex}
\begin{tabular}{|c|r|r|r|r|r|r|}
\hline
Graph & \multicolumn{3}{|c|}{With cost model} & \multicolumn{3}{|c|}{Without cost model} \\

\cline{2-7}

fragments    & $|\tilde{D}|$ & $|E_{\tilde{D}}|$ & $\timec$(s) & $|D|$ & $|E_{D}|$ & $\timec$(s) \\

\hline\hline
CO      &  32.1   & 537.8   & 0.1     &49.8   &549.4      &0.1\\ \hline
FL      &  39.5   & 689.3   & 0.2      &61.7   &705.7     &0.2\\ \hline
CA      &  51.3   & 1,021.9 & 0.4     &78.4   &1,045.1    &0.4\\ \hline
E-US    &  71.1   & 1617.1  & 0.9     &107.0  &1651.8     &0.8\\ \hline
W-US    &  68.9   & 1,541.6 & 0.9     &104.6  &1,576.8    &0.9\\ \hline
C-US    &  116.4  & 3,251.3 & 4.0    &169.4  &3,329.8     &3.9\\ \hline
US      &  124.9  & 3,584.3 & 4.9    &183.1  &3,673.4     &4.8\\ \hline
\end{tabular}
\vspace{-4ex}
\end{scriptsize}
\end{center}
\vspace{-4ex}
\end{table}

\begin{table}[t!]
\vspace{4ex}
\label{tab-exp2-super-graph}
\caption{Sizes of \super graphs}
\begin{center}
\begin{scriptsize}
\vspace{-2ex}
\begin{tabular}{|c|r|r|r|r|r|r|r|}
\hline
$\G$ & CO     & FL     & CA     & E-US     & W-US      & C-US      & US        \\
\hline
$|\V_c|/|V|$         &\hspace{-1ex}3.9\%\hspace{-2ex}&\hspace{-2ex} 3.0\%   \hspace{-2ex}&\hspace{-2ex}  2.9\%  &\hspace{-2ex}  2.8\%  &\hspace{-2ex}  2.1\%   &\hspace{-2ex} 2.3\%   &\hspace{-2ex} 1.8\%     \\
\hline
$|\E_c|/|E|$        &\hspace{-1ex}14.5\%\hspace{-2ex}&\hspace{-2ex} 10.9\%  \hspace{-2ex}&\hspace{-2ex} 12.7\%  &\hspace{-2ex} 14.2\% &\hspace{-2ex} 10.3\%  &\hspace{-2ex} 14.5\%  &\hspace{-2ex} 12.0\% \\

\hline
$|\V|/|V|$         &\hspace{-1ex}3.9\%\hspace{-2ex}&\hspace{-2ex} 3.0\%   \hspace{-2ex}&\hspace{-2ex}  2.9\%  &\hspace{-2ex}  2.8\%  &\hspace{-2ex}  2.1\%   &\hspace{-2ex} 2.3\%   &\hspace{-2ex} 1.8\%     \\
\hline
$|\E|/|E|$        &\hspace{-1ex}14.8\%\hspace{-2ex}&\hspace{-2ex} 11.1\%  \hspace{-2ex}&\hspace{-2ex} 13.0\%  &\hspace{-2ex} 14.5\% &\hspace{-2ex} 10.5\%  &\hspace{-2ex} 14.5\%  &\hspace{-2ex} 12.3\%\\
\hline
\end{tabular}
\vspace{-4ex}
\end{scriptsize}
\end{center}
\vspace{-3.5ex}
\end{table}

\etitle{Exp-3: Impacts of hybrid landmark covers}.
In the third set of experiments, using the graph fragments generated at Exp-2, we evaluated (1) the average number of nodes and edges enforced by the hybrid landmarks covers  with or without the cost model, and (2) their average efficiency on a single fragment. The results are reported in Table~5.

The results tell us that the usage of the cost model both reduces the number of landmarks and enforced edges, moreover, it only incurs little extra time cost.

We also report the \super graphs in Table~6.
The \super graphs $\G$ are quite small, typically have 2--4\% nodes and 10--15\% edges compared with the original graphs $G(V, E)$.
Using hybrid landmark covers with the cost model, the \super graphs $\G(\V_c, \E_c)$ further reduce 0.2--0.3\% edges. This justified the effectiveness of agents and graph partitions, and the introduction of the cost model for hybrid landmark covers.

\begin{figure*}[tb!]
\begin{center}
\subfigure[{\scriptsize Vary \# of nodes}]{\label{fig-exp3-extra-space-cost-1}
\vspace{-4ex}
\includegraphics[scale=0.436]{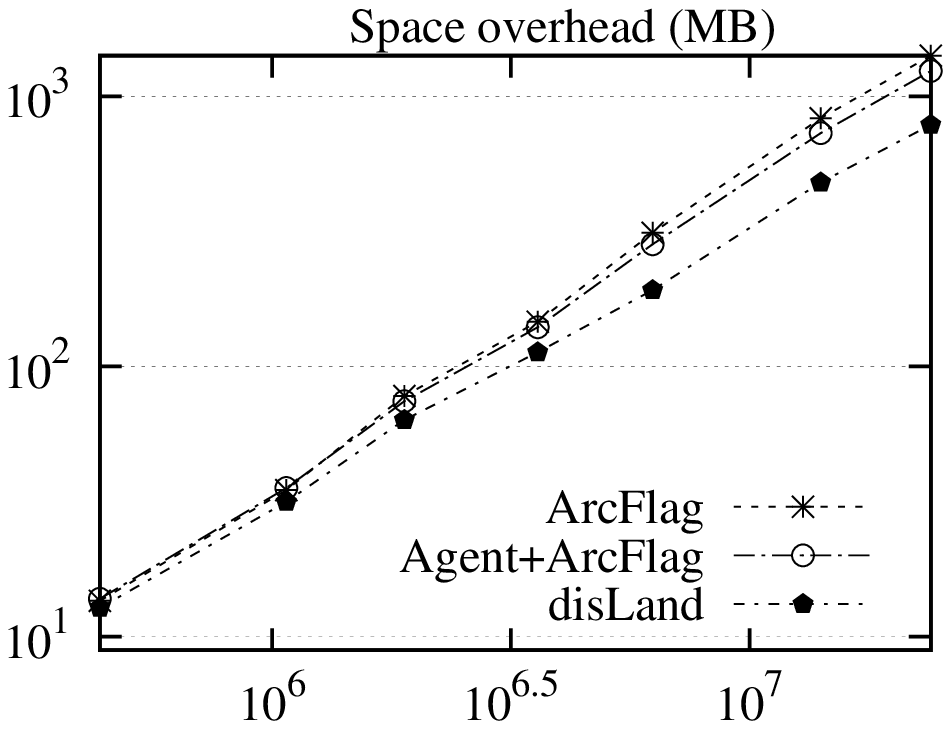}}
\hspace{-5ex}
\subfigure[{\scriptsize Vary \# of nodes}]{\label{fig-exp3-extra-space-cost-2}
\vspace{-4ex}
\includegraphics[scale=0.436]{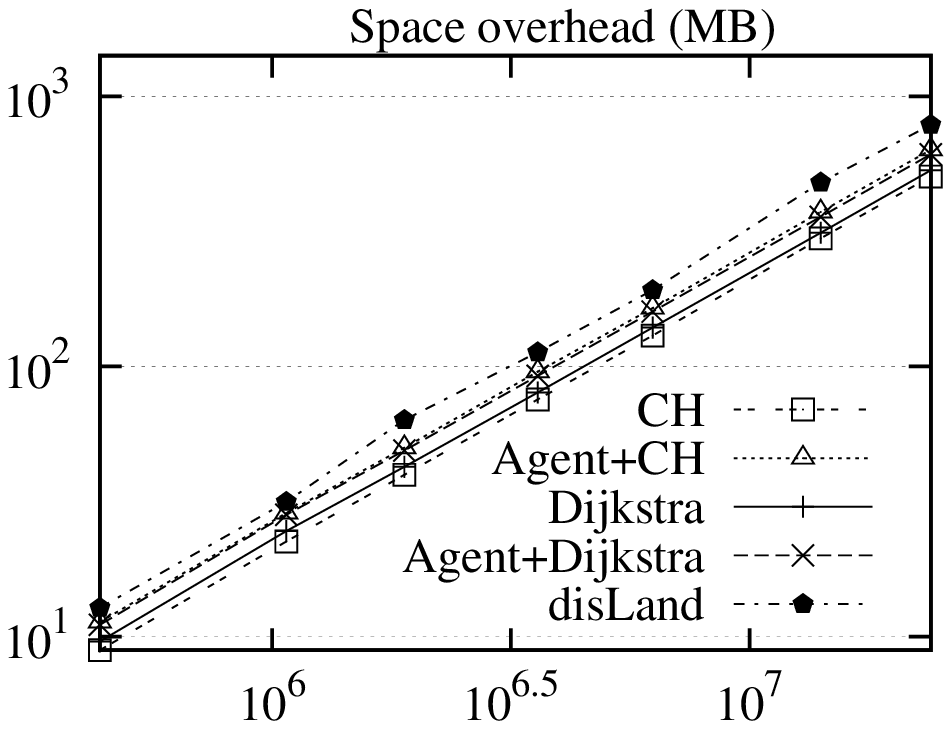}}
\hspace{-5ex}
\subfigure[{\scriptsize Vary \# of nodes}]{\label{fig-exp3-preprocessing-time-1}
\vspace{-4ex}
\includegraphics[scale=0.436]{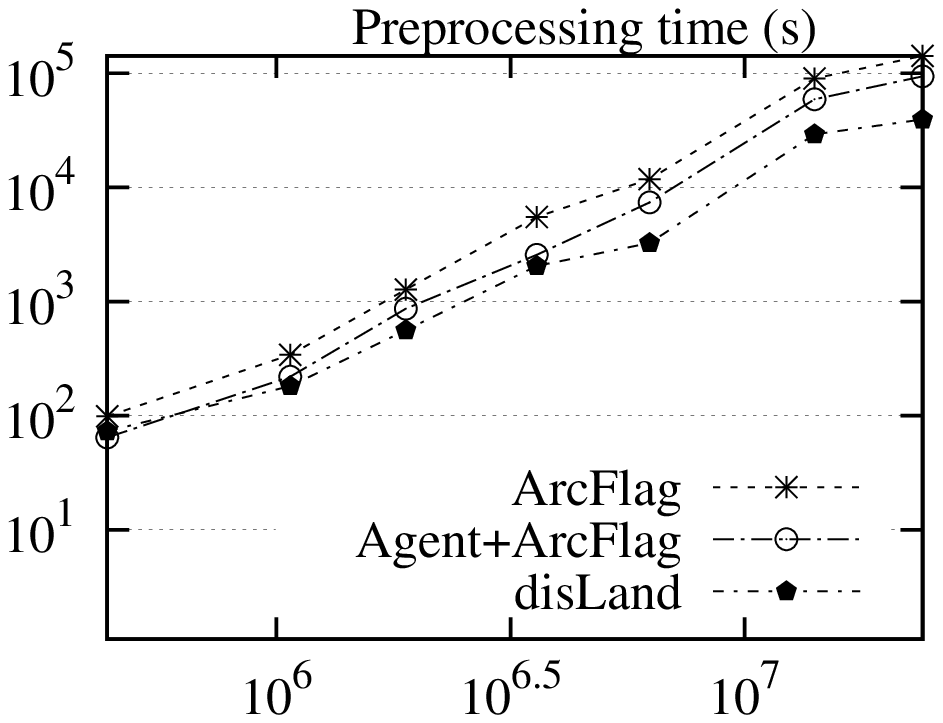}}
\hspace{-5ex}
\subfigure[{\scriptsize Vary \# of nodes}]{\label{fig-exp3-preprocessing-time-2}
\vspace{-4ex}
\includegraphics[scale=0.436]{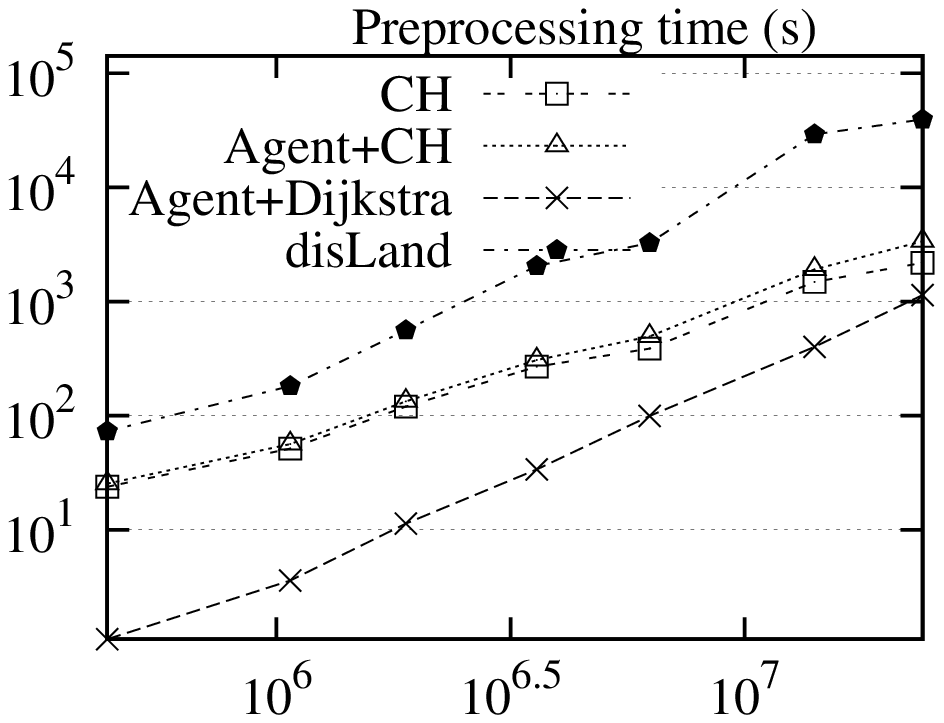}}
\end{center}
\vspace{-3ex}
\caption{Space overhead and preprocessing time} \label{fig-exp-space-time-cost}
\vspace{-2ex}
\end{figure*}

\begin{figure*}[tb!]
\begin{center}
\subfigure[{\scriptsize $Q_1$}]{\label{fig-exp4-varySize-Q1}
\includegraphics[scale=0.42]{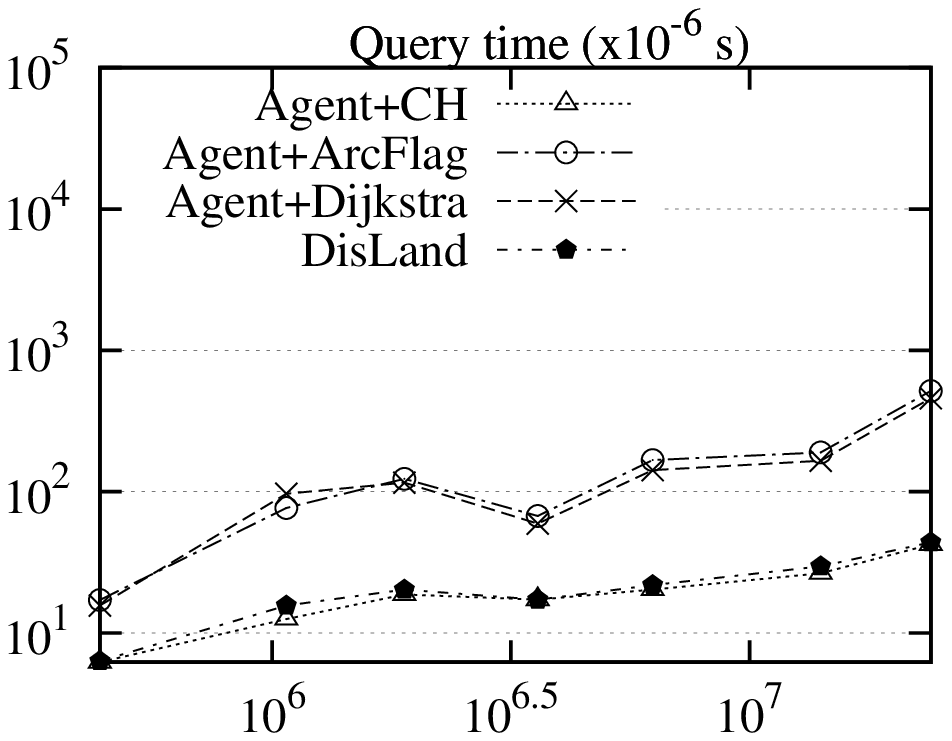}}
\hspace{-4ex}
\subfigure[{\scriptsize $Q_2$}]{\label{fig-exp4-varySize-Q2}
\includegraphics[scale=0.42]{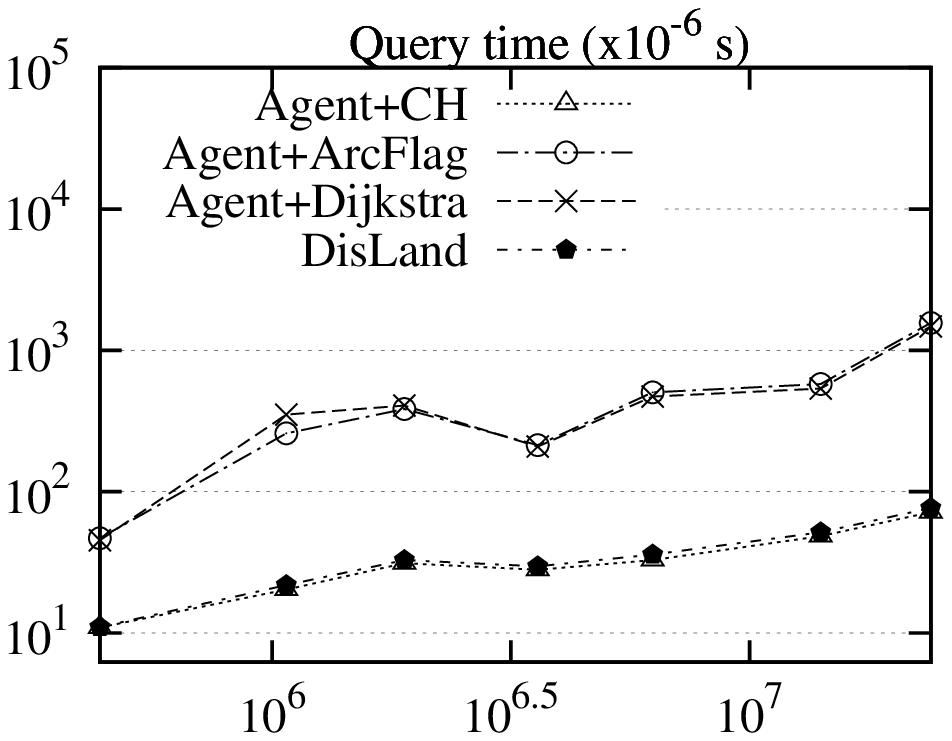}}
\hspace{-4ex}
\subfigure[{\scriptsize $Q_3$}]{\label{fig-exp4-varySize-Q3}
\includegraphics[scale=0.42]{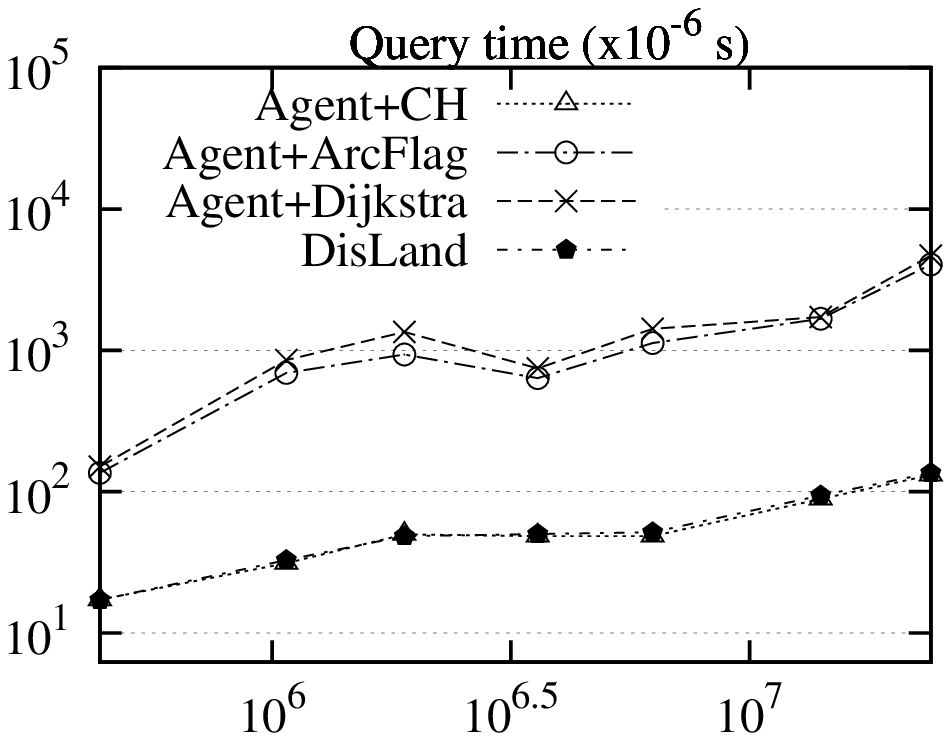}}
\hspace{-4ex}
\subfigure[{\scriptsize $Q_4$}]{\label{fig-exp4-varySize-Q4}
\includegraphics[scale=0.42]{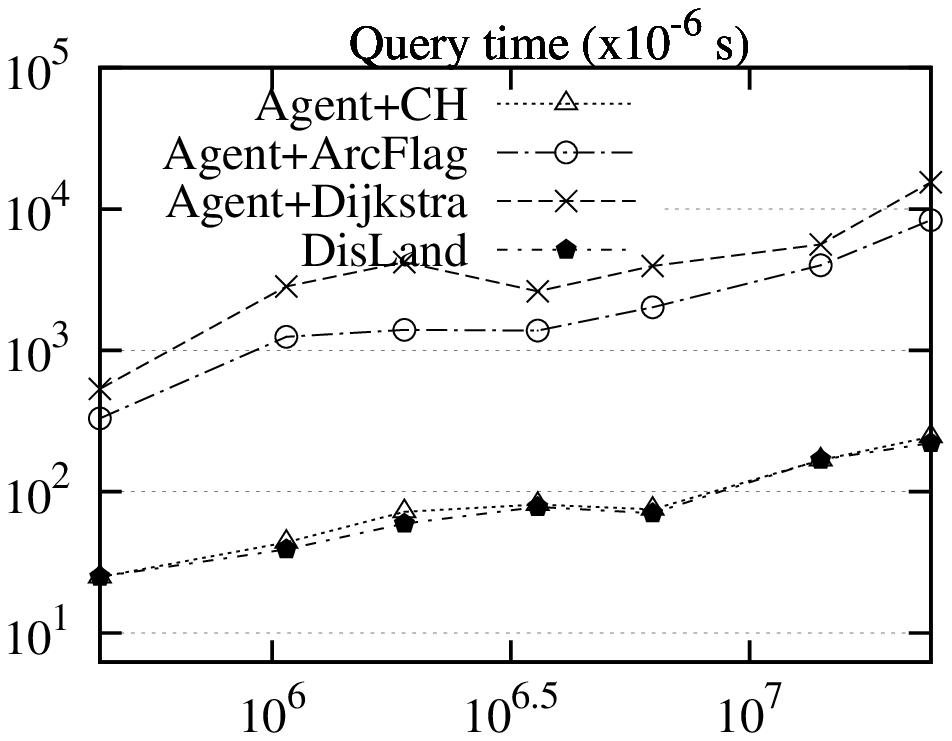}}
\hspace{-4ex}\vspace{-1ex}
\subfigure[{\scriptsize $Q_5$}]{\label{fig-exp4-varySize-Q5}
\includegraphics[scale=0.42]{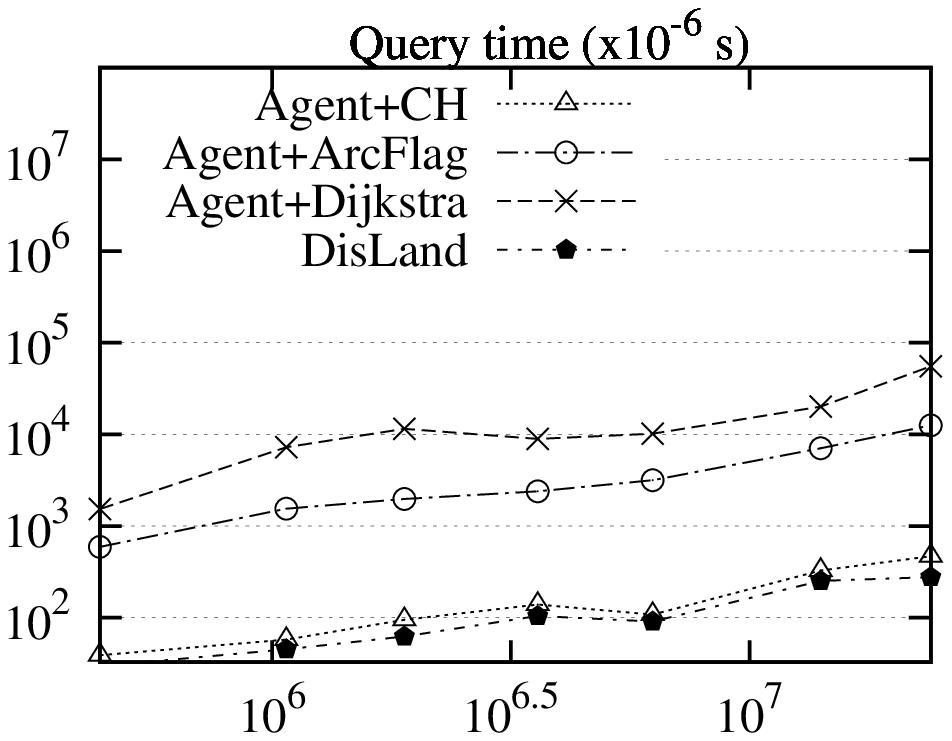}}
\hspace{-4ex}\vspace{-1ex}
\subfigure[{\scriptsize $Q_6$}]{\label{fig-exp4-varySize-Q6}
\includegraphics[scale=0.42]{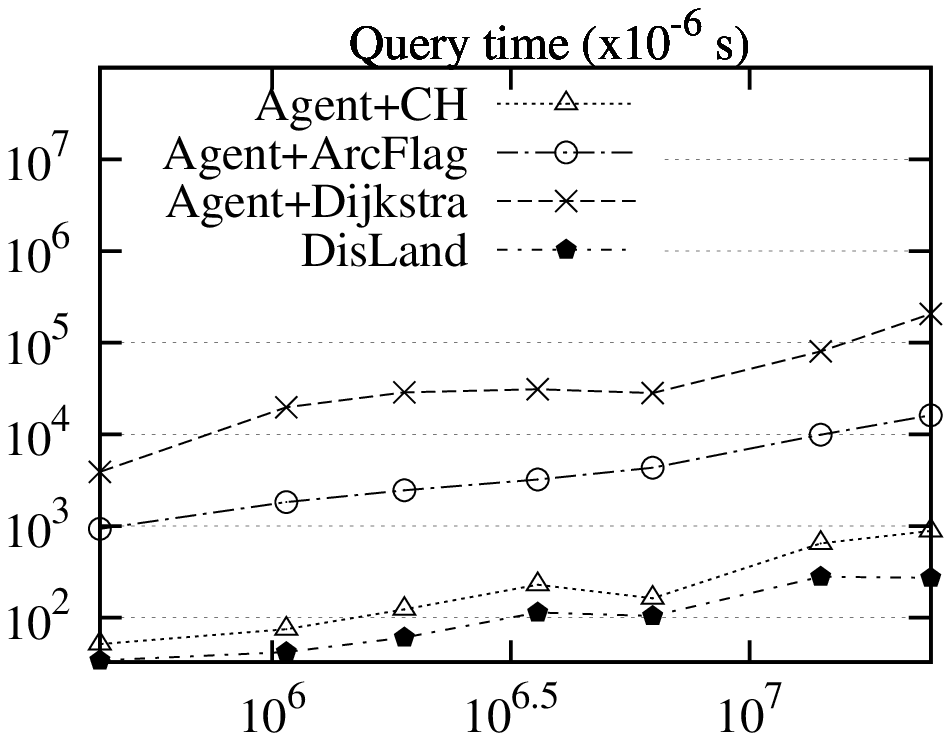}}
\hspace{-4ex}\vspace{-1ex}
\subfigure[{\scriptsize $Q_7$}]{\label{fig-exp4-varySize-Q7}
\includegraphics[scale=0.42]{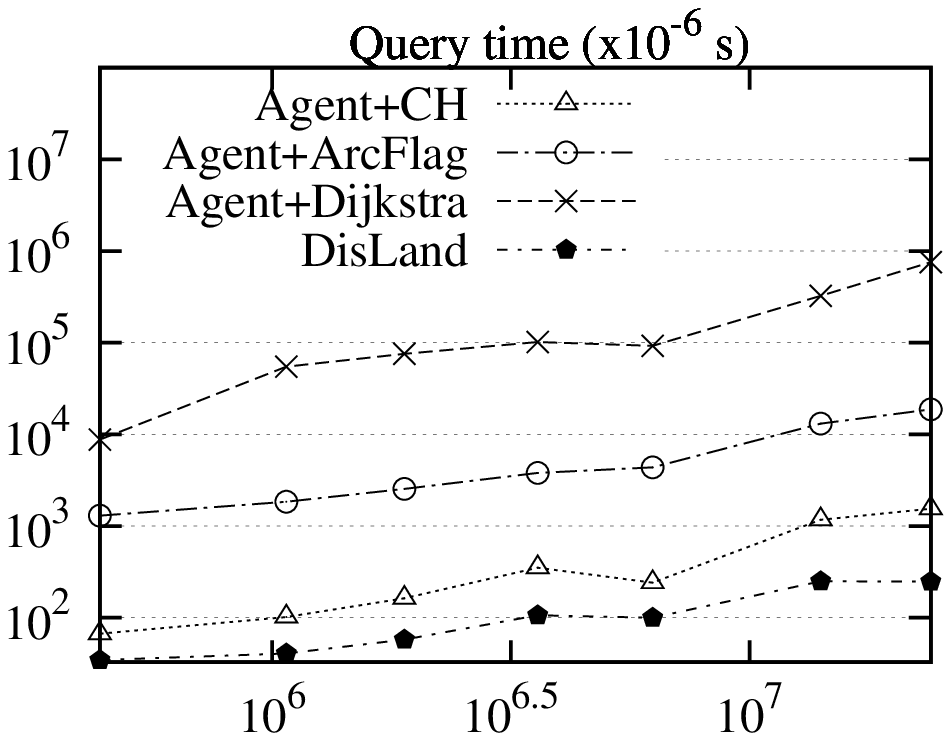}}
\hspace{-4ex}\vspace{-1ex}
\subfigure[{\scriptsize $Q_8$}]{\label{fig-exp4-varySize-Q8}
\includegraphics[scale=0.42]{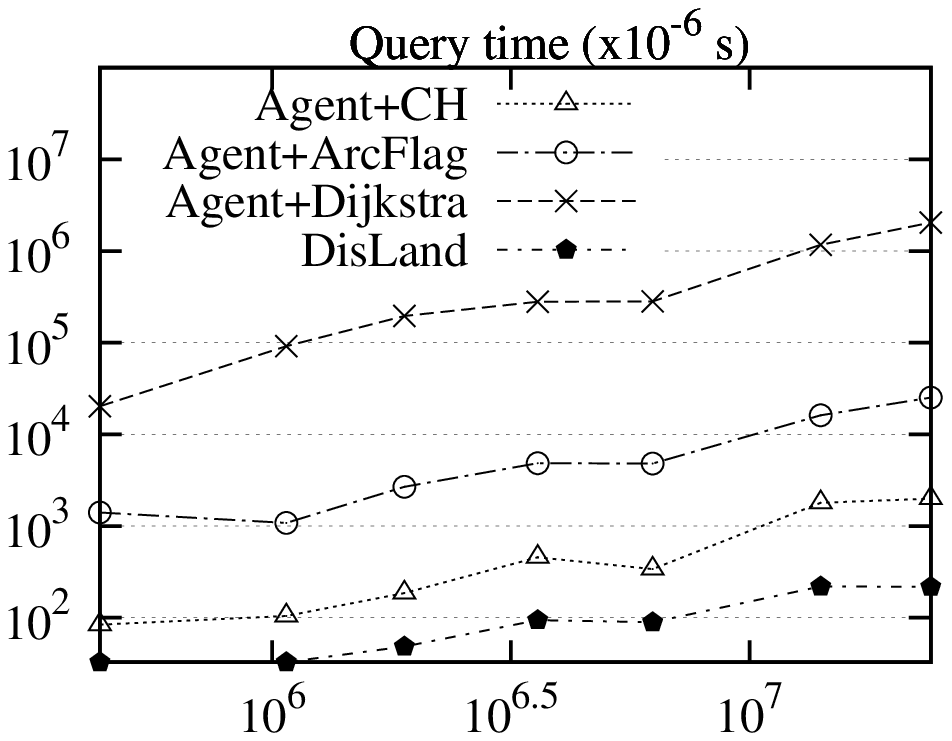}}
\end{center}
\vspace{0ex}
\caption{Performance evaluation \wrt graph sizes} \label{fig-exp4-varySize}
\vspace{-2ex}
\end{figure*}

\begin{figure*}[tb!]
\begin{center}
\hspace{10ex}
\subfigure[{\scriptsize CO}]{\label{fig-exp4-varyQ-CO}
\includegraphics[scale=0.42]{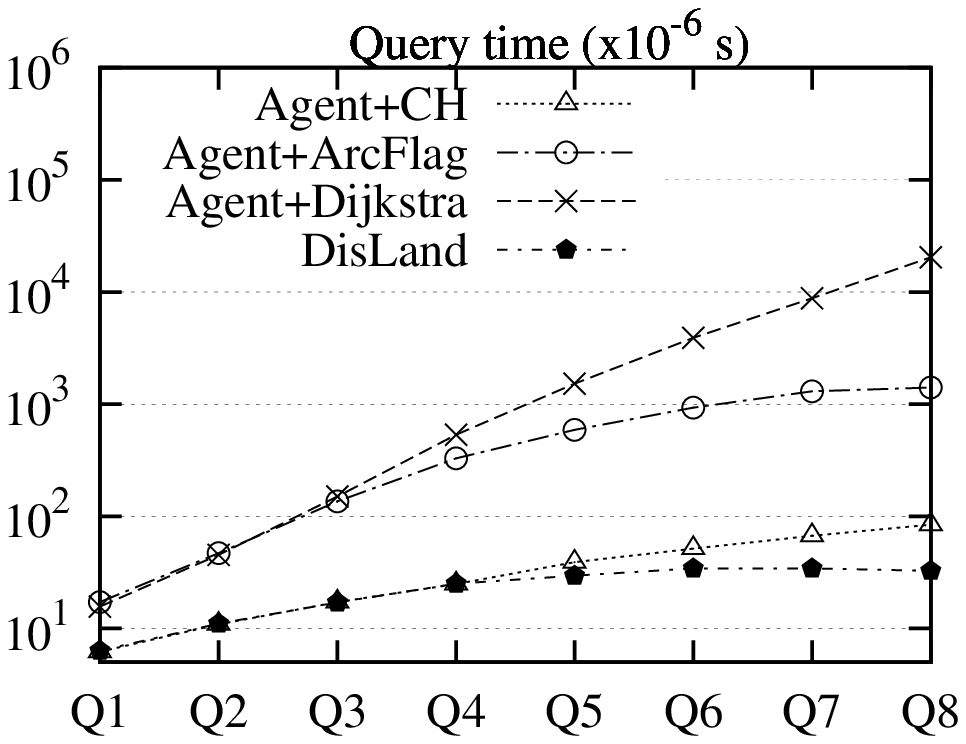}}
\hspace{-4ex}
\subfigure[{\scriptsize FL}]{\label{fig-exp4-varyQ-FL}
\includegraphics[scale=0.42]{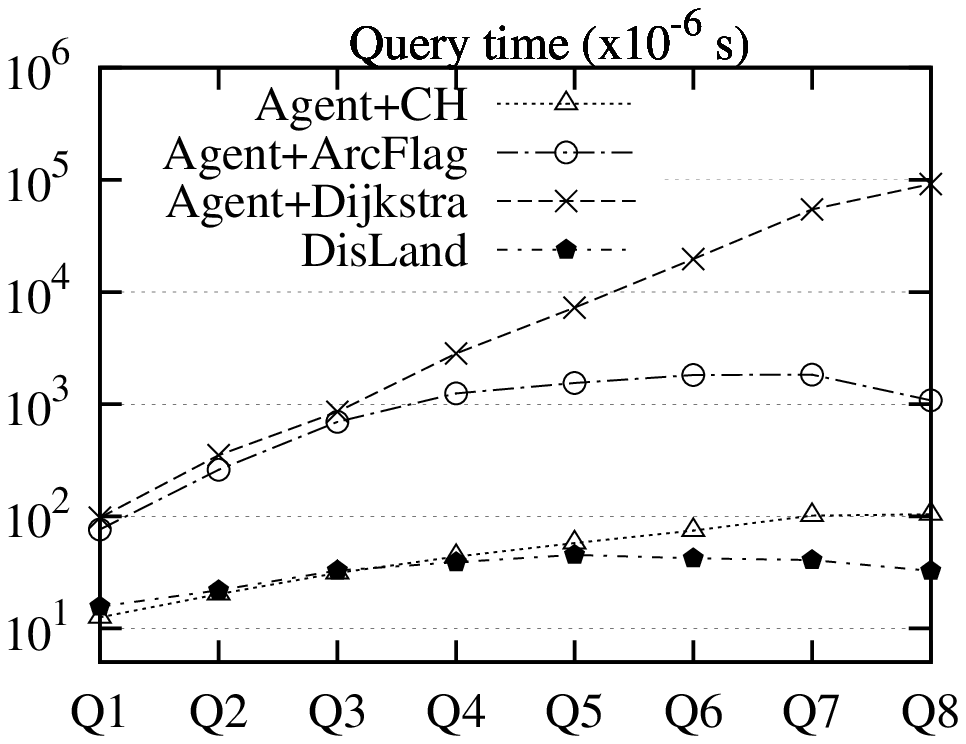}}
\hspace{-4ex}
\subfigure[{\scriptsize CA}]{\label{fig-exp4-varyQ-CA}
\includegraphics[scale=0.42]{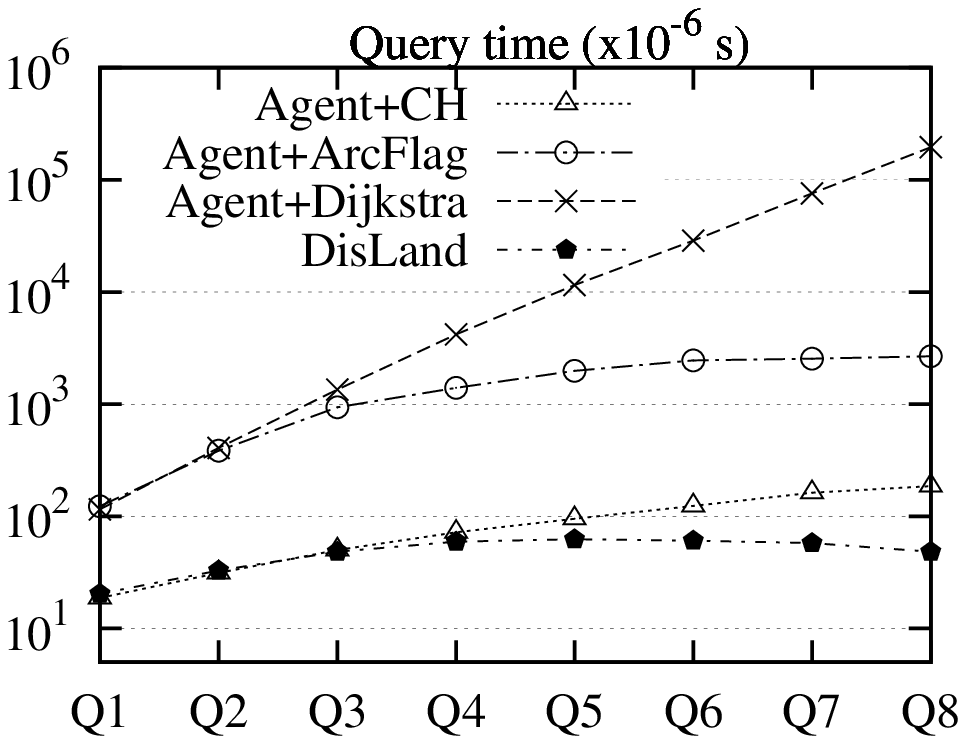}}
\hspace{10ex}\vspace{-1.5ex}

\subfigure[{\scriptsize E-US}]{\label{fig-exp4-varyQ-E-US}
\includegraphics[scale=0.42]{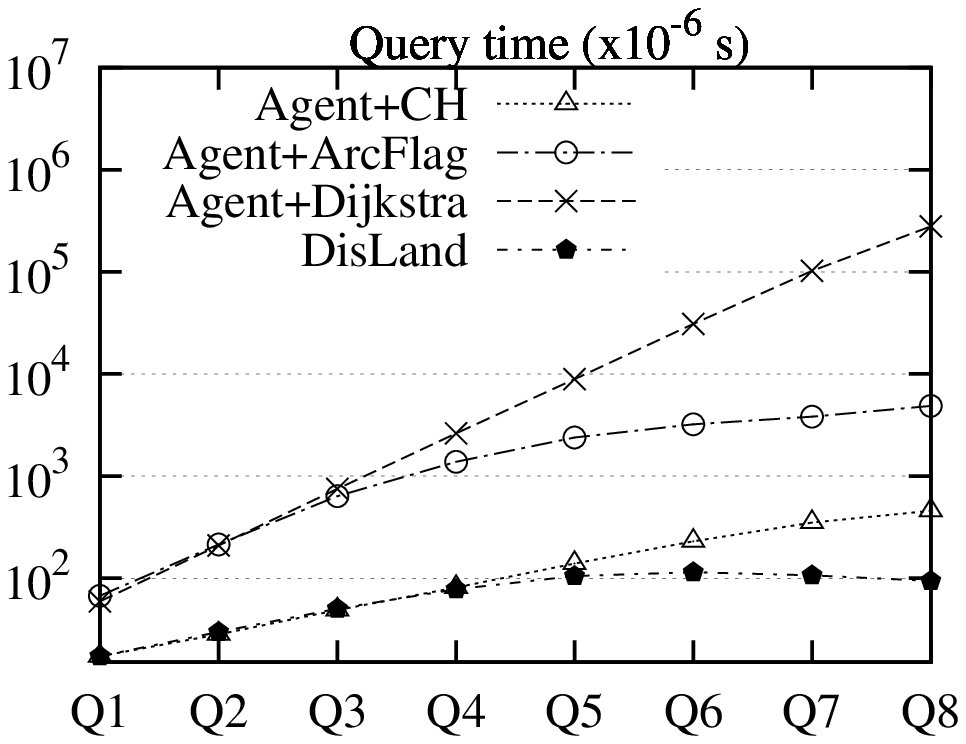}}
\hspace{-4ex}\vspace{-1.5ex}
\subfigure[{\scriptsize W-US}]{\label{fig-exp4-varyQ-ME-W-US}
\includegraphics[scale=0.42]{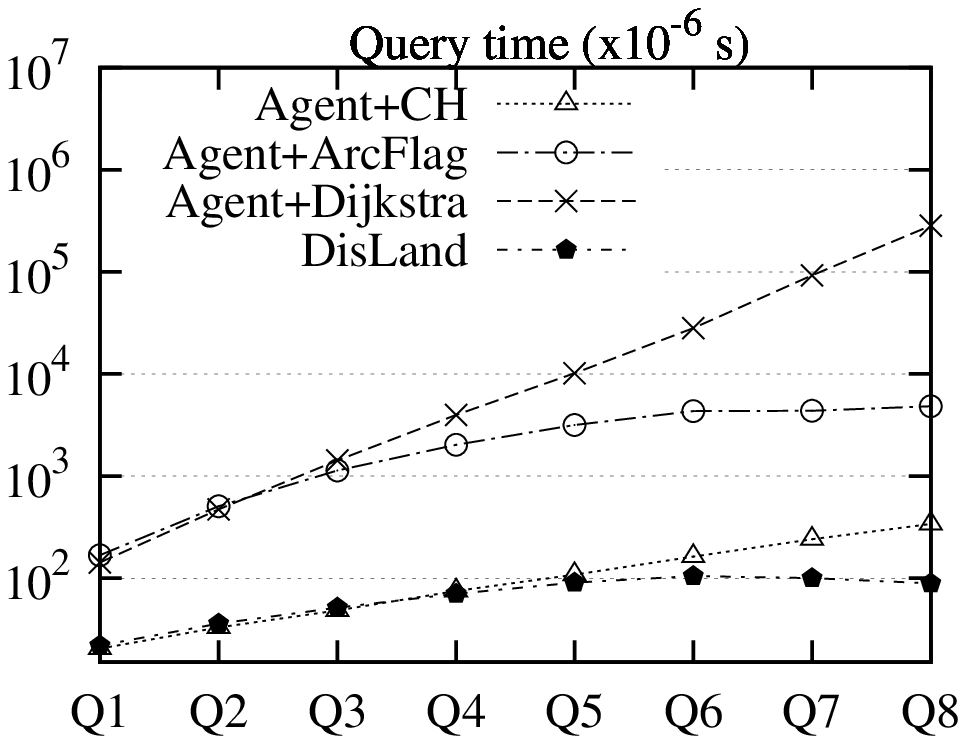}}
\hspace{-4ex}\vspace{-1.5ex}
\subfigure[{\scriptsize C-US}]{\label{fig-exp4-varyQ-C-US}
\includegraphics[scale=0.42]{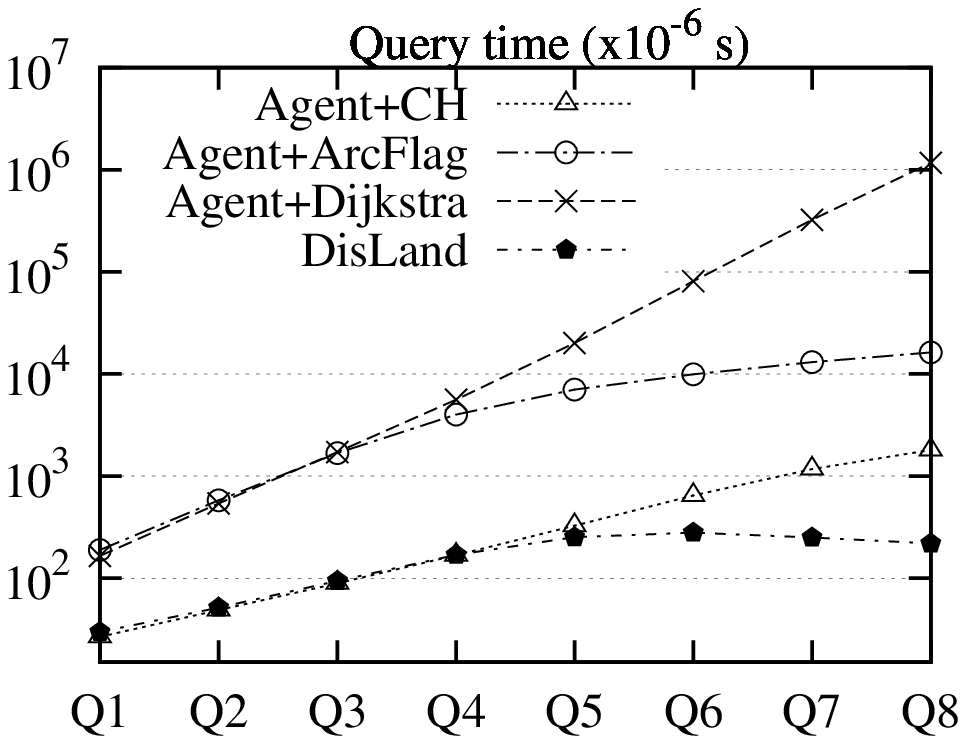}}
\hspace{-4ex}\vspace{-1.5ex}
\subfigure[{\scriptsize US}]{\label{fig-exp4-varyQ-US}
\includegraphics[scale=0.42]{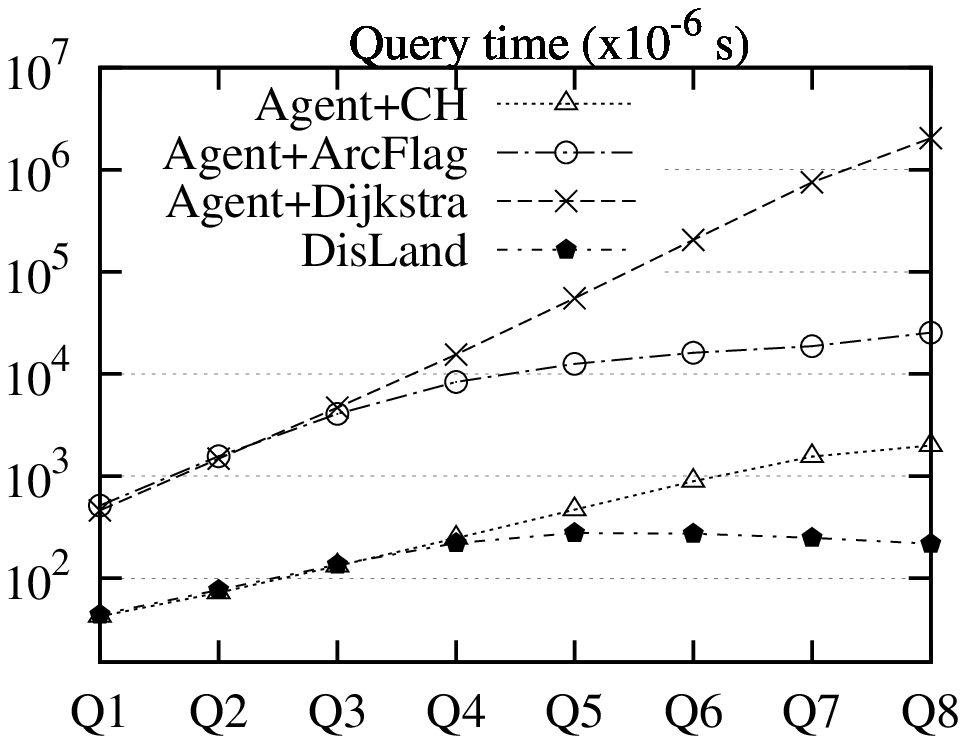}}
\end{center}
\vspace{2ex}
\caption{Performance evaluation \wrt distance queries} \label{fig-exp4-varyQ}
\vspace{-4ex}
\end{figure*}

\etitle{Exp-4: Preprocessing time and space overhead}. In the fourth set of experiments, we tested the space cost and preprocessing time of Dijkstra, Agent + Dijkstra,
\ch, Agents + \ch, \arcflag, Agents + \arcflag, and \disland.  For \disland, we did a second level partition on the \super graphs into $k$ fragments, determined as follows: $k = \lfloor \frac{m}{1000} \rfloor \cdot 100$ if $m$ > $1000$, and $k = \lfloor \frac{m}{100} \rfloor \cdot 10$, otherwise, where $m$ is the number of fragments of the shrink graphs, shown in Table~4. \arcflag called \metis to partition the graphs into $k$ fragments as well. The results are reported in Figure~\ref{fig-exp-space-time-cost}.

The results tell us that (1) the space cost follows the order: \arcflag $>$ Agents + \arcflag $>$ \disland $>$ Agents + \ch $>$ Agent + Dijkstra $>$ Dijkstra $>$ \ch;
and (2) the preprocessing time follows the order: \arcflag $>$ Agents + \arcflag $>$ \disland $>$ Agents + \ch $>$ \ch $>$ Agent + Dijkstra.
In particular, \ch even uses less space than the original graphs, and \disland uses about $1/2$ time extra space, while Agent + \arcflag and \arcflag use $1.66$ and $1.24$  times extra space, respectively. While \ch and \disland could finish the preprocessing in less than $0.5$ and $11$ hours, repectively, it took Agent + \arcflag and \arcflag $26$ and $40$ hours, respectively.  Thus all approaches, except \arcflag and Agents + \arcflag, produce auxiliary structures with a small space cost and in a reasonably fast way.

\eat{
We will add arc flags to edges in our DisLand method, arc flag method and arc flag with Agents. When adding arc flags in DisLand method, only super graph were considered. We partitioned the super graph into $k$ fragments added a $k$-bit wide label to each edge in the super graph. The fragments number $k$ is determined as follows. Let $m$ be the number of fragments we have partitioned in the former phase. If $m$ $>$ $1000$, then $k = \lfloor \frac{m}{1000} \rfloor \cdot 100$. Otherwise, $k = \lfloor \frac{m}{100} \rfloor \cdot 10$. For each boundary node in fragment $i \in [1, k]$, we set the $i$th bit true in the label on every edge in the shortest path DAG in the reverse super graph.

Arc flag method and Arc flag with Agents method partitioned the graph into same number of fragments as we did in DisLand method.

time: \ch vs. \disland vs. \disland with \ch

The time cost of each techniques is reported in table 9. IO operation are not considered.

space: \ch vs. \disland vs. \disland with \ch

The size of index generated by each techniques is reported in table 10. We compute the size of index by combining the edge size and label size. Let $n$ denote the number of nodes in the graph and $m$ the number of all edges, including shortcut connecting a node in DRA and its agent, super edges in Super Graph. $E_{SUPER}$ is a set of edges in Super Graph. $b$ is the size of a label, which is equal to $ (\lfloor f/32 \rfloor + 1) \cdot 4$ byte. The size of this index is $(|E_{Super}| \cdot b + n+(m \cdot 2) \cdot 4)$ byte.
}

\etitle{Exp-5: Efficiency of shortest distance queries}. In the last set of experiments, using the $8$ sets $Q_1, \ldots, Q_8$ of distance queries, we tested the efficiency of Dijkstra, Agent + Dijkstra, \ch, Agents + \ch, \arcflag, Agents + \arcflag, and \disland on the $7$ datasets with corresponding generated auxiliary structures. The results are reported in Figures~\ref{fig-exp4-varySize} and~\ref{fig-exp4-varyQ}. As for all algorithms, their counterparts with agents were always faster, we omitted their running time for clarity.

The results tell us that (1) all algorithms scale well \wrt the graph sizes and \wrt the distance queries, and (2) the efficiency of the algorithms follows the order: \disland, Agent + \ch $>$ Agent + \arcflag $>$ Agent + Dijkstra.  For the distance queries ($Q_1, \ldots, Q_4$) with relative close distance node pairs, the running time of \disland and Agent + \ch is comparable. However, for the distance queries ($Q_5, \ldots, Q_8$) with relative long distance node pairs, \disland is apparently faster than Agent + \ch. Indeed, for $Q_8$ on the US dataset, \disland is $14,540.1$, $9,430.2$, $134.9$, $116.5$ , $9.4$ and $9.1$ times faster than Dijkstra, Agent + Dijkstra, \arcflag, Agent + \arcflag, \ch, and Agent + \ch, respectively.

\vspace{-0.5ex}
\stitle{Summary}. From these experimental results, we find the following. (1) \disland scales well on large road graphs, \eg it takes only $0.28 \times 10^{-3}$ seconds on graphs with $2.4$ $\times$ $10^7$ nodes and $5.7$ $\times$ $10^7$ edges. (2) Agents and their \dras are a light-weight preprocessing technique, which benefits almost all shortest distance algorithms. (3) Agents, graph partitions and hybrid landmark covers together provide a good solution to produce small \super graphs, which typically have 2--4\% nodes and 10--15\% edges compared with the original graphs. (4)  \disland produces auxiliary structures with a small space cost (about $1/2$ of the input graphs), and their preprocessing could be finished in a reasonably fast way. (5) \disland provides a good solution for shortest distance query answering, especially for far node pairs on large graphs. For $Q_8$ on the US dataset, it is even $9.1$ times faster than Agent + \ch, where \ch is the best approach without using extra information, \eg longitude and latitude, tested in~\cite{WuXDCZZ12}. Finally, (6) hybrid landmark covers play a central role that makes our proposed techniques (\eg agents and graph partitions) and the existing techniques (\eg\ \ch and \arcflag) seamlessly integrate into a unified framework -- \disland.

\section{Conclusion}
\label{sec-con}

We have studied how to apply distance landmarks for fast exact shortest distance query answering on large weighted undirected road graphs. To our knowledge, we are among the first to settle this problem.
We have shown that the direct application of distance landmarks is impractical due to their high space and time cost. To rectify these problems, we have proposed: hybrid landmark covers, agents and \dras, bounded graph partitions, \super graphs
and framework \disland.  We have also verified,
both analytically and experimentally, that  hybrid landmark covers, together with these techniques, significantly
improve efficiency of shortest distance queries.

Several topics are targeted for future work.
%
%
We are to extend our techniques for other types real-life datasets that could be modeled as weighted undirected graphs, \eg social networks.
We are also to explore the possibility of applying distance landmarks for other classes of graph queries, \eg reachability.

\eat{
\section{Conclusion}
\label{sec-con}

We have studied how to apply distance landmarks for fast exact shortest distance query answering on large weighted undirected road graphs. To our knowledge, we are among the first to settle this problem.
We have shown that the direct application of distance landmarks is impractical due to their high space and time cost. To rectify these problems, we have proposed: hybrid landmark covers, agents and \dras, bounded graph partitions, \super graphs
and framework \disland.  We have also verified,
both analytically and experimentally, that  hybrid landmark covers, together with these techniques, significantly
improve efficiency of shortest distance queries.

Several topics are targeted for future work.
%
%
We are to extend our techniques for other types real-life datasets that could be modeled as weighted undirected graphs, \eg social networks.
We are also to explore the possibility of applying distance landmarks for other classes of graph queries, \eg reachability.

\stitle{Acknowledgments}. Shuai is supported in part by  NGFR 973 grant
2011CB302602 and NSFC grants 90818028 and 60903149.
}


\balance
\vspace{-1ex}
\bibliographystyle{abbrv}
\vspace{2ex}
\begin{small}
\bibliography{paper}

\begin{thebibliography}{10}

\bibitem{AbrahamDGW11}
I.~Abraham, D.~Delling, A.~V. Goldberg, and R.~F.~F. Werneck.
\newblock A hub-based labeling algorithm for shortest paths in road networks.
\newblock In {\em SEA}, 2011.

\bibitem{BroderKMRRSTW00}
A.~Z. Broder, R.~Kumar, F.~Maghoul, P.~Raghavan, S.~Rajagopalan, R.~Stata,
  A.~Tomkins, and J.~L. Wiener.
\newblock Graph structure in the web.
\newblock {\em Computer Networks}, 33(1-6):309--320, 2000.

\bibitem{ch-algorithm}
{CH}.
\newblock {\sl http://algo2.iti.kit.edu/english/routeplanning.php}.

\bibitem{ChanL07}
E.~P.~F. Chan and H.~Lim.
\newblock Optimization and evaluation of shortest path queries.
\newblock {\em VLDB J.}, 16(3):343--369, 2007.

\bibitem{ChengKCC12}
J.~Cheng, Y.~Ke, S.~Chu, and C.~Cheng.
\newblock Efficient processing of distance queries in large graphs: a vertex
  cover approach.
\newblock In {\em SIGMOD}, 2012.

\bibitem{CormenLRS01}
T.~H. Cormen, C.~E. Leiserson, R.~L. Rivest, and C.~Stein.
\newblock {\em Introduction to Algorithms}.
\newblock The MIT Press, 2001.

\bibitem{Dijkstra59}
E.~W. Dijkstra.
\newblock A note on two problems in connexion with graphs.
\newblock {\em Numerische Mathematik}, 1:269--271, 1959.

\bibitem{dimacs-datasets}
{DIMACS}.
\newblock {\sl http://www.dis.uniroma1.it/challenge9}.

\bibitem{corr-abs-MF}
M.~Franceschet.
\newblock Collaboration in computer science: A network science approach.
\newblock {\em JASIST}, 62(10):1992--2012, 2011.

\bibitem{FredmanT84}
M.~L. Fredman and R.~E. Tarjan.
\newblock Fibonacci heaps and their uses in improved network optimization
  algorithms.
\newblock In {\em FOCS}, 1984.

\bibitem{full}
{Full version}.
\newblock http://mashuai.buaa.edu.cn/full.pdf.

\bibitem{GaJo79}
M.~Garey and D.~Johnson.
\newblock {\em Computers and Intractability: A Guide to the Theory of
  {NP}-Completeness}.
\newblock W. H. Freeman and Company, 1979.

\bibitem{GeisbergerSSD08}
R.~Geisberger, P.~Sanders, D.~Schultes, and D.~Delling.
\newblock Contraction hierarchies: Faster and simpler hierarchical routing in
  road networks.
\newblock In {\em WEA}, 2008.

\bibitem{GoldbergH05}
A.~V. Goldberg and C.~Harrelson.
\newblock Computing the shortest path: {\it A*} search meets graph theory.
\newblock In {\em SODA}, 2005.

\bibitem{HopcroftT73-cut-nodes}
J.~E. Hopcroft and R.~E. Tarjan.
\newblock Efficient algorithms for graph manipulation [h] (algorithm 447).
\newblock {\em Commun. ACM}, 16(6):372--378, 1973.

\bibitem{Karypis98}
G.~Karypis and V.~Kumar.
\newblock A fast and high quality multilevel scheme for partitioning irregular
  graphs.
\newblock {\em SISC}, 20(1):359--392, 1998.

\bibitem{kl70}
B.~W. Kernighan and S.~Lin.
\newblock An efficientheuristic procedure for partitioning graphs.
\newblock {\em Bell System Technical Journal}, 49(1):13--21, 1970.

\bibitem{LappasLT09}
T.~Lappas, K.~Liu, and E.~Terzi.
\newblock Finding a team of experts in social networks.
\newblock In {\em KDD}, 2009.

\bibitem{LeskovecLDM08}
J.~Leskovec, K.~J. Lang, A.~Dasgupta, and M.~W. Mahoney.
\newblock Statistical properties of community structure in large social and
  information networks.
\newblock In {\em WWW}, 2008.

\bibitem{LubyR89}
M.~Luby and P.~Ragde.
\newblock A bidirectional shortest-path algorithm with good average-case
  behavior.
\newblock {\em Algorithmica}, 4(4):551--567, 1989.

\bibitem{metis}
{Metis}.
\newblock {\sl http://glaros.dtc.umn.edu/gkhome/views/metis}.

\bibitem{MohringSSWW05}
R.~H. M{\"o}hring, H.~Schilling, B.~Sch{\"u}tz, D.~Wagner, and T.~Willhalm.
\newblock Partitioning graphs to speedup {D}ijkstra's algorithm.
\newblock {\em ACM Journal of EA}, 11, 2006.

\bibitem{MozesS12}
S.~Mozes and C.~Sommer.
\newblock Exact distance oracles for planar graphs.
\newblock In {\em SODA}, 2012.

\bibitem{PotamiasBCG09}
M.~Potamias, F.~Bonchi, C.~Castillo, and A.~Gionis.
\newblock Fast shortest path distance estimation in large networks.
\newblock In {\em CIKM}, 2009.

\bibitem{SandersS05}
P.~Sanders and D.~Schultes.
\newblock Highway hierarchies hasten exact shortest path queries.
\newblock In {\em ESA}, 2005.

\bibitem{SankaranarayananS10}
J.~Sankaranarayanan and H.~Samet.
\newblock Query processing using distance oracles for spatial networks.
\newblock {\em TKDE}, 22(8):1158--1175, 2010.

\bibitem{SankaranarayananSA09}
J.~Sankaranarayanan, H.~Samet, and H.~Alborzi.
\newblock Path oracles for spatial networks.
\newblock {\em PVLDB}, 2(1), 2009.

\bibitem{SarmaGNP10}
A.~D. Sarma, S.~Gollapudi, M.~Najork, and R.~Panigrahy.
\newblock A sketch-based distance oracle for web-scale graphs.
\newblock In {\em WSDM}, 2010.

\bibitem{SaundersT07}
S.~Saunders and T.~Takaoka.
\newblock Solving shortest paths efficiently on nearly acyclic directed graphs.
\newblock {\em TCS}, 370(1-3):94--109, 2007.

\bibitem{ThorupZ05}
M.~Thorup and U.~Zwick.
\newblock Approximate distance oracles.
\newblock {\em J. ACM}, 52(1):1--24, 2005.

\bibitem{approx03}
V.~V. Vazirani.
\newblock {\em Approximation Algorithms}.
\newblock Springer, 2003.

\bibitem{WagnerW07}
D.~Wagner and T.~Willhalm.
\newblock Speed-up techniques for shortest-path computations.
\newblock In {\em STACS}, 2007.

\bibitem{Wei10}
F.~Wei.
\newblock Tedi: efficient shortest path query answering on graphs.
\newblock In {\em SIGMOD}, 2010.

\bibitem{WuXDCZZ12}
L.~Wu, X.~Xiao, D.~Deng, G.~Cong, A.~D. Zhu, and S.~Zhou.
\newblock Shortest path and distance queries on road networks: An experimental
  evaluation.
\newblock {\em PVLDB}, 5(5), 2012.

\bibitem{YangYZK12}
S.~Yang, X.~Yan, B.~Zong, and A.~Khan.
\newblock Towards effective partition management for large graphs.
\newblock In {\em SIGMOD}, 2012.

\end{thebibliography}
\end{small}

\end{document}